\documentclass[a4paper,11pt]{article}
\usepackage{titlesec}
\titleformat{\paragraph}[runin]{\normalfont\itshape}{\theparagraph.}{.3em}{}[.]\titlespacing{\paragraph}{0pt}{1ex plus .1ex minus .2ex}{.5em}
\usepackage{amsmath,amssymb,bm}
\usepackage[T1]{fontenc}
\usepackage{mathtools}
\usepackage{dsfont}
\pdfoutput=1
\usepackage[english]{babel}
\usepackage[letterpaper, hmargin=1in, top=1in, bottom=1.2in, footskip=0.6in]{geometry}
\usepackage{graphicx} % Allows including images
\usepackage{booktabs} % Allows the use of \toprule, \midrule and \bottomrule in tables
\usepackage{color}
\definecolor{aquamarine}{rgb}{0.5, 1.0, 0.83}
\definecolor{ao(english)}{rgb}{0.0, 0.5, 0.0}
\definecolor{armygreen}{rgb}{0.29, 0.33, 0.13}
\definecolor{awesome}{rgb}{1.0, 0.13, 0.32}
\definecolor{ballblue}{rgb}{0.13, 0.67, 0.8}
\definecolor{bittersweet}{rgb}{1.0, 0.44, 0.37}
\definecolor{blue}{rgb}{0.0, 0.0, 1.0}
\definecolor{brinkpink}{rgb}{0.98, 0.38, 0.5}
\definecolor{ballblue}{rgb}{0.13, 0.67, 0.8}
\definecolor{brightturquoise}{rgb}{0.03, 0.91, 0.87}
\definecolor{blue-green}{rgb}{0.0, 0.87, 0.87}
\definecolor{caribbeangreen}{rgb}{0.0, 0.8, 0.6}
\definecolor{cyan}{rgb}{0.0, 1.0, 1.0}
\definecolor{amber(sae/ece)}{rgb}{1.0, 0.49, 0.0}
\graphicspath{ {images/} }
\definecolor{vdarkred}{rgb}{0.6,0,0.2}

\author{J\"urg Fr\"ohlich$^1$}

\title{Irreversibility and the Arrow of Time}

\begin{document}

\maketitle
\begin{abstract}
Within the general formalism of quantum theory \textit{irreversibility} and the \textit{arrow of time} 
in the evolution of various physical systems are studied.
Irreversible behavior often manifests itself in the guise of \textit{``entropy production.''} 
This motivates me to begin this paper with a brief review of quantum-mechanical 
entropy, a subject that \textit{Elliott Lieb} has made outstanding contributions to, followed by an 
enumeration of examples of irreversible behavior and of an arrow of time analyzed in later sections.
Subsequently, a derivation of the laws of thermodynamics from (quantum) statistical mechanics,
and, in particular, of the \textit{Second Law of thermodynamics}, in the forms given to it by 
\textit{Clausius} and \textit{Carnot}, is presented.
In a third part, results on \textit{diffusive (Brownian) motion} of a quantum particle interacting with a 
quasi-free quantum-mechanical heat bath are reviewed. This is followed by an outline of a 
theory of \textit{friction} by emission of Cherenkov radiation of sound waves in a system 
consisting of a particle moving through a Bose-Einstein condensate and interacting with it.
In what may be the most important section of this paper, the \textit{fundamental arrow of time} 
inherent in \textit{Quantum Mechanics} is discussed. 
\end{abstract}

\section{Introductory Remarks}\label{IR}
As \textit{Elliott Lieb} likes to say, it is often at least as rewarding to discover and\,/\,or explore
new and sometimes unexpected consequences of known Laws of Nature as it is to discover new such Laws. 
In this paper, some ideas and results on \textit{irreversibility} and the \textit{arrow of time} in the behavior 
of physical systems, derived from mostly well established theory, are reviewed.

The \textit{arrow of time} is a short hand for the basic dichotomy of \textit{past} and \textit{future}, the past 
consisting of facts or \textit{actualities}, and the future consisting of uncertainties and \textit{potentialities.} 
In a time-reversal invariant, deterministic world inhabited by an infinitely clever mind who is able to compute 
all processes, past and future -- definitely not our world! -- there would be no arrow of time. 

In this paper, a physical system is said to exhibit an \textit{arrow of time} if complete knowledge of its past
states (its actualities) does \textit{not} suffice to predict its future states with certainty; and\,/\,or if knowledge of 
its \textit{present} state does \textit{not} contain enough information to enable one to reconstruct
\textit{past} states of the system, in case one failed to observe them, i.e., if information about the past -- 
if not recorded -- is dissipating.\footnote{Aristotle put it as follows: \textit{``Indeed, it is 
evident that the mere passage of time itself is destructive rather than generative ..., because change 
is primarily a `passing away'.''}}
For example, knowing the state of a Brownian walker at time $t$ does not enable one to determine 
its states at times $<t$ (in case one did not record them), nor its states at times $>t$; see Sect.~\ref{QBM}. 
In quantum mechanics, knowing the exact state at time $t$ of a system featuring \textit{events} 
does usually \textit{not} enable one to reconstruct its states at times $<t$, let alone to predict 
future states, with certainty; see Sect.~\ref{ETH}.\footnote{\textit{``Every experiment destroys some 
of the knowledge of the system which was obtained by previous experiments.''} (Werner Heisenberg)}

By \textit{irreversibility} I mean the following: A dynamical process happening in a physical system is 
called \textit{irreversible} if, for all practical purposes, it is impossible to run the process \textit{backwards}. 
For example, motion with friction, studied in Sect.~\ref{Friction}, if run backwards, would become continuously 
accelerated motion, which one does not observe; or then, a steady flow of heat energy from a colder 
reservoir to a hotter reservoir is never observed; see Sect.~\ref{Fund Law}.

This paper is based on a lecture about irreversibility and the arrow of time given by the author 
at various institutions, during the past seven years.\footnote{It was first presented, on invitation kindly
extended to me by A. Alekseev, as a \textit{Mirimanoff lecture} at the University of Geneva, in 2015.} 
It addresses a theme that has been discussed by numerous authors (see \cite{Lebowitz} and 
references given there), and one might expect that not much can be said that is really new. 
The results described on the following pages have grown out of collaborations of the author 
with quite a number of colleagues and friends that have extended over at least two decades; 
and I think that some of our results have clarified some important aspects of the general topic
alluded to in the title and abstract. 

In this paper, I focus on a quantum-mechanical description of the physical systems to be considered; but I am 
confident that many of the results discussed in Sections \ref{Fund Law} - \ref{Friction} can also be derived -- 
mutatis mutandis -- within classical physics. My survey is intended for physicists; mathematical precision 
is not its main priority. However, where I state results in the form of theorems the reader can be confident 
that there are mathematically precise statements and proofs thereof in the literature referred to in the text.

It is with great pleasure that I dedicate this paper to an admired colleague, mentor and friend of many years, 
\textit{Elliott H. Lieb}, on the occasion of his 90$^{th}$ birthday -- with all best wishes!

\section{Relative Entropy}\label{entropy}
In this section we recapitulate the definition of quantum-mechanical entropy \cite{LR} and some of its properties used 
afterwards. Of particular importance for the purposes of this paper is \textit{relative entropy}.

\noindent Let $\Omega$ be a density matrix acting on a Hilbert space $\mathcal{H}$; (i.e., $\Omega$ is a quantum-mechanical ``state''). 
The \textit{von Neumann entropy} of $\Omega$ is defined by
\begin{equation}\label{Entropy}
S(\Omega) := - \text{Tr} \big(\Omega \cdot \text{ln} \Omega\big)
\end{equation}
It has the following properties:
\begin{itemize}
 \item[\quad1.]{$S(\Omega) \geq 0,\, \forall \Omega,$ with ``$=$'' only if $\Omega$ is pure, i.e., a rank-1 orthogonal projection.}
 \item[\quad2.]{$S(\cdot)$ is strictly concave.}
 \item[\quad3.]{$S(\cdot)$ is subadditive and strongly subadditive; see \cite{LR}}
\end{itemize}
Statement 1 is obvious. Statement 2 and subadditivity of entropy follow from

\textit{Klein’s Inequality:} 

Let $f$ be a real-valued, strictly convex function on the real line, 
and let $A$ and $B$ be bounded, self-adjoint, trace-class operators on $\mathcal{H}$. Then
\begin{equation}\label{Klein}
\text{Tr}\big(f (B)\big) \geq \text{Tr} \big(f(A)\big) + \text{Tr} \big(f'(A) · (B-A)\big), 
\end{equation}
with ``$=$'' only if $A=B$. To prove concavity and subadditivity of $S(\cdot)$ one chooses $f(x)=x \cdot \text{ln}x$.

\textit{Proof of Inequality} \eqref{Klein}:
This proof is simple enough to be presented here. Let $\{\psi_ j\}_{j=0}^{\infty}$ be a complete orthonormal 
system (CONS) of eigenvectors of $B$ corresponding to eigenvalues $\beta_j, j=0,1,2,\dots$ Let $\psi$ be an
arbitrary unit vector in $\mathcal{H}$, and set $c_j :=\langle \psi_j , \psi\rangle, \forall j$. Then convexity of $f$ implies the
following two inequalities:
\begin{align}\label{lowerbd}
\langle \psi, f(B)\psi \rangle &= \sum_{j}\vert c_j \vert^{2} f(\beta_j) \geq
f\big(\sum_{j} \vert c_j \vert^{2}\beta_j\big) = f\big(\langle \psi, B \psi\rangle \big)\\
f\big(\langle \psi, B \psi\rangle \big)&\geq f\big((\langle \psi, A \psi\rangle \big) + f'\big(\langle \psi, A \psi\rangle\big)\cdot 
\langle \psi, (B-A)\psi \rangle\,. \label{derivative}
\end{align}
If $\psi$ is an eigenvector of $A$ then the right side of \eqref{derivative} is given by
\begin{equation}\label{eigenvector}
=\langle \psi,\big[f(A)+ f'(A)\cdot(B-A)\big]\psi \rangle\hspace{0.45cm}
\end{equation}
Eq. \eqref{Klein} follows by summing inequalities \eqref{lowerbd} and \eqref{eigenvector} over a CONS of eigenvectors of A.

Let $\Sigma$ and $\Omega$ be two arbitrary density matrices on 
$\mathcal{H}$. The \textit{relative entropy} of $\Sigma$ with respect to $\Omega$ is defined by
\begin{equation}\label{rel entropy}
S(\Sigma\,\Vert \Omega):= Tr(\Sigma\,(\text{ln}\Sigma - \text{ln}\Omega)),
\end{equation}
assuming that ker($\Omega$) $\subseteq \text{ker}\Sigma$. 

Crucial properties of $S(\Sigma\,\Vert \Omega)$ are:
\begin{itemize}
\item{\textit{Positivity:}
\begin{equation}\label{positivity}
S(\Sigma\,\Vert \Omega) \geq 0, \quad \text{as follows from inequality \eqref{Klein}} 
\end{equation}}
\item{\textit{Convexity:} $$S(\Sigma\,\Vert \Omega) \, \text{ is jointly convex in } \, \Sigma \, \text{ and }\,  \Omega.$$}
\end{itemize}
Joint convexity can be derived from the following inequality: Let $\mathcal{T}$ be a trace-preserving, completely positive map on the convex set of density matrices on $\mathcal{H}$. Then
\begin{equation*}
S(\Sigma\,\Vert \Omega) \geq S(\mathcal{T}(\Sigma)\,\Vert \mathcal{T}(\Omega))\,.
\end{equation*}
\textit{Exercise:} Show that this inequality, due to \textit{Lindblad} and \textit{Uhlmann} \cite{LU, Sutter}, implies \textit{Strong Subadditivity}, first established by \textit{Lieb} and \textit{Ruskai} \cite{LR}:
$$S(\Omega_{12})+S(\Omega_{23})-S(\Omega_{123})-S(\Omega_{2})\geq 0\,,$$
where $\Omega_{123}$ is a density matrix on $\mathcal{H}_{1}\otimes \mathcal{H}_{2}\otimes \mathcal{H}_{3}$, 
and the density matrices $\Omega_{12}\,,$
... are obtained by taking partial traces. A well known application of strong subadditivity is to proving existence
of the thermodynamic limit of specific entropy. 

For the purposes of this paper, inequality \eqref{positivity} is particularly important. It will be used to derive 
the \textit{$2^{nd}$ Law of Thermodynamics} from quantum statistical mechanics; see Sect.~\ref{Fund Law}.

I have learned the neat proof of Klein's Inequality and the right way of introducing the $2^{nd}$ 
\textit{Law of thermodynamics} from my teacher \textit{Res Jost} (see \cite{Jost}). Incidentally, he also warned some of us that, at a party, one should never start a conversation 
about 
\begin{itemize}
\item{irreversibility and the arrow of time;}
\item{the foundations and the ``interpretation'' of Quantum Mechanics;}
\item{religious faith.}
\end{itemize}
For, most people mistakenly believe that they have some solid understanding of these topics, 
and they get surprisingly emotional when one tells them that they might actually be wrong. 
(I have made the experiment myself.)

There appears to be much confusion -- even among grown-up theoretical physicists -- about \textit{irreversibility} 
and the origin of \textit{time's arrow}. For example, most people mistakenly believe that all \textit{fundamental 
laws of Nature} treat past and future in a symmetric way (i.e., do not distinguish between past and future). 
And, in the author's modest opinion, there is quite an enormous confusion about the foundations 
of \textit{Quan- tum Mechanics} and its deeper meaning!

In this paper, I attempt to uncover causes of the \textit{arrow of time} and of \textit{irreversibility} in physics, 
in particular in quantum physics (touching on its foundations). 

The following examples of physical systems exhibiting irreversible behavior and\,/\, or an arow of time will be considered.
\begin{itemize}
\item{Irreversibility: A macroscopic physical system prepared in an unlikely initial state (e.g., a small subsystem coupled 
to some macroscopic reservoirs at different temperatures and\,/\,or chemical potentials; the Universe) -- see 
Sects.~\ref{Fund Law} and \ref{UCA}.}
\item{Arrow of time: A small subsystem, e.g., a quantum-mechanical particle, evolving under the influence of noise 
from its environment, such as a heat bath -- this is the subject matter of Sect.~\ref{QBM}, where a theory of 
Quantum Brownian Motion is sketched.} 
\item{Irreversibility: A physical system, such as a particle coupled to the electromagnetic field or to a Bose gas exhibting 
Bose-Einstein condensation, leaking energy and ``information'' into massless modes that propagate to $\infty$ and, 
hence, are not accessible to observation, anymore -- this forms the contents of Sect.~\ref{Friction}, where the 
phenomenon of \textit{friction by emission of Cherenkov radiation} is analyzed.}
\item{Arrow of time: Isolated physical systems featuring \textit{``events,''} as described by Quantum Mechanics -- this fundamental example is treated in Sect.~\ref{ETH}.}
\end{itemize}

\section{The Second Law of Thermodynamics -- Clausius and Carnot} \label{Fund Law}

Elliott Lieb has had a successful and widely noted collaboration with J.~Yngvason on the foundations of
thermodynamics \cite{LYng}. It may therefore be appropriate to review some recent and not so recent results 
concerning a derivation of the laws of thermodynamics from (quantum) statistical mechanics and their 
implications concerning the irreversibility of certain natural processes.

\vspace{0.2cm}\noindent
{\bf{I. \,Preliminaries on thermodynamics from the point of view of quantum 
statistical mechanics}}

We begin this section by recalling some facts in thermodynamics that can be derived from quantum statistical
mechanics.

Let $R$ be a macroscopically large thermal reservoir (or heat bath) whose state space is a separable Hilbert 
space, denoted by $\mathcal{H}_R$, and whose Hamiltonian is denoted by $H_R$. In order for our arguments 
to be meaningful mathematically, we will initiallly suppose that reservoirs are finitely extended in space, i.e., 
are confined to a compact region (e.g., a rectangle), $\Lambda$, in physical space, $\mathbb{E}^{3}$. 
Subsequently the thermodynamic limit, $\Lambda \nearrow \mathbb{E}^{3}$, will most often be taken.
But we will not always make it explicit whether the reservoirs are assumed to be finitely extended or to fill all of 
physical space, and we will not discuss conditions that guarantee the existence of the thermodynamic limit. 
These matters have been widely discussed in the literature; see, e.g., \cite{Ruelle}. We suppose that every family
of thermal reservoirs (indexed by spatial regions $\Lambda$ they are confined to) considered in the following can 
be described by a net, $\big(\mathcal{A}(O)\big)_{O\subset \mathbb{E}^{3}}$, of local algebras of operators 
representing physical quantities of the reservoirs, where $O$ is an arbitrary bounded open set in physical 
space, and $\mathcal{A}(O_1)\subseteqq \mathcal{A}(O_2)$ whenever 
$O_1 \subset O_2$; see \cite{Br-Rob}. We may then define
\begin{equation}\label{quasi-loc alg}
\mathcal{A}:= \bigvee_{O\subset \mathbb{E}^{3}} \mathcal{A}(O),
\end{equation}
(or $\mathcal{A}=$ closure of the right side in the operator norm).
We assume that, in the thermodynamic limit, a thermal reservoir $R$ obeys the \textit{Zero$^{th}$ Law} 
of thermodynamics: Initial states of $R$ within a ``large class of states'' -- when restricted to a local algebra
$\mathcal{A}(O)$ -- approach states that are \textit{indistinguishable} from thermal equilibrium 
states, restricted to the algeba $\mathcal{A}(O)$, as time $t$ tends to $\infty$, for an arbitrary bounded region $O$. 
(A somewhat more precise formulation of the \textit{Zero$^{th}$ Law} can be found at the end of this section.)

Let $C$ be a ``small system'' coupled to $R$. The state space of $C$ is denoted by $\mathcal{H}_C$; 
the one of the total system, $S=R \vee C$, is then chosen to be
$\mathcal{H}_S:=\mathcal{H}_R \otimes \mathcal{H}_C$. The Hamiltonian of $S$ is given by
$$H(t):= H_R\vert_{\mathcal{H}_R} \otimes \mathbf{1}_{\mathcal{H}_C} + H_C(t),$$
where $H_C(t)$ is a possibly time-dependent bounded \textit{quasi-local operator} on $\mathcal{H}_S$. 
For the purposes of this paper quasi-local operators are defined to be operators contained in the algebra
$\mathcal{A} \otimes B(\mathcal{H}_C)$, with $B(\mathcal{H}_C)$ the algebra of bounded operators 
on $\mathcal{H}_C$, and $\mathcal{A}$ as in \eqref{quasi-loc alg}. We assume that $H(t) \rightarrow H_{\infty}$, 
as $t\rightarrow \infty$, in a sense made more precise in Eq.~\eqref{Liouville bound} below.
Let $\big\{U(t,s)\,\vert\, t, s \text{ in } \mathbb{R}\big\}$ be the unitary propagator generated by the time-dependent
Hamiltonians $H(t)$, i.e.,
\begin{align}\label{prop}
\frac{\partial}{\partial t}U(t,s)&= -iH(t)\, U(t,s)\,, \quad \forall t, s,\\
U(t,u)\cdot U(u,s)& = U(t,s)\,, \,\,\,\, U(t,t) = \mathbf{1}, \,\,\,\forall\, t,u,s\,.\nonumber
\end{align}
(We will always use units such that $\hbar=1$.) See, e.g., \cite{R-S} for results concerning \eqref{prop}. 
General states of $S$ are described by \textit{density matrices}, $\Omega, \Sigma, \dots$, on 
$\mathcal{H}_S$. Let 
$$\Omega_t = U(t,s) \Omega U(s,t),$$
 be the true state of $S$ at time $t$, where
$\Omega$ is the initial state of $S$ prepared at time $s$.
The equation of motion of $\Omega_t$ is given by
\begin{equation}\label{Liouville}
\frac{\partial}{\partial t} \Omega_t = -i[H(t), \Omega_t] \equiv -i \mathcal{L}(t)\big(\Omega_t\big)\,,
\end{equation}
where $\mathcal{L}(t)$ is the so-called \textit{``Liouvillian''} of the system; (the Liouvillians $\mathcal{L}(t), t\in \mathbb{R},$ 
can be viewed as self-adjoint operators on $\mathcal{H}_S \otimes \mathcal{H}_S$).

We also define the \textit{instantaneous thermal equilibrium state} at inverse temperature $\beta=1/k_B T$ associated with the 
Hamiltonian $H(t)$ to be given by

$$\Omega^{\beta}_t := Z_t(\beta)^{-1} \text{exp}[-\beta\, H(t)], \quad \text{with}\quad Z_t(\beta)= \text{Tr}\big(\text{exp}[-\beta\, H(t)]\big),$$
(assuming that the region $\Lambda$ the system is confined to is bounded). It is convenient to use the notation
\begin{equation}\label{states}
\omega_t (A):= \text{Tr}\big(\Omega_t \cdot A\big), \qquad \omega_t^{\beta}(A):= \text{Tr}\big(\Omega_t^{\beta} \cdot A\big), 
\end{equation}
where $A\in \mathcal{A}\otimes B(\mathcal{H}_C)$ is an arbitrary quasi-local operator. 
We will usually assume that, for every quasi-local $A$, the thermodynamic limit, $\Lambda \nearrow \mathbb{E}^{3}$, 
of the expectation values $\omega_t(A)$ exists (at least in the sense of convergent subsequences), 
\textit{uniformly} in time $t$ in bounded intervals of the time axis. (We will not indicate the dependence 
of these expectation values on $\Lambda$.) 

\vspace{0.2cm}\noindent \textit{Definition of some thermodynamic quantities:}
\begin{itemize}
\item{Internal energy: $U(t):= \omega_{t}\big(H(t)\big)$ \, \text{ and } \,  $U_{C}(t):= \omega_{t}\big(H_{C}(t)\big)$\,.}
\item{Reservoir power: $\mathcal{P}(t)\equiv \mathcal{P}_{R}(t)\equiv \dot{Q}(t) := \frac{d}{dt} \omega_{t}\big(H_R\big)
 = -i\omega_{t}\big([H_R, H_C(t)]\big)$, where $[H_R, H_C(t)]$ is a quasi-local operator; (so that the right side may have a
well defined thermodynamic limit).}
\end{itemize}
We then have that
\begin{equation}\label{bilance}
\dot{U}(t)= \dot{Q}(t) + \dot{U}_{C}(t) = \omega_{t}\big(\dot{H}_C(t)\big)\,.
\end{equation}
For a time-dependent function $f(t), t\in \mathbb{R},$ we define
$$\Delta f \equiv \Delta_{s}^{t}f:= f(t) - f(s) = \int_{s}^{t} \dot{f}(\tau) d\tau.$$
Integrating Eq.~\eqref{bilance} in time, we find that
\begin{equation}\label{first law}
\Delta U_{C}=\Delta Q^{\swarrow} + \Delta W,
\end{equation}
where $\Delta Q^{\swarrow} = -\Delta Q$ is the heat energy absorbed by the subsystem $C$, and 
$$\Delta W \equiv \Delta_{s}^{t}W:= \int_{s}^{t}\omega_{\tau}\big(\dot{H}_{C}(\tau)\big)\,d\tau$$
 is the work done on $C$ between time $s$ and time $t$. Eq.~\eqref{first law} is the \textit{First Law} of thermodynamics.

Next, we review an important example of irreversible behavior, namely the phenomenon of \textit{return to equilibrium} 
(see \cite{L-Rob, JP, BFS, FM}), which expresses a remarkable stability property of thermal equilibrium against 
local perturbations. We assume that the Liouvillians of $S$ (see Eq.~\eqref{Liouville}) have the following properties:
\begin{itemize}
\item[\quad(i)]{The Liouvillians $\mathcal{L}_t, t \in \mathbb{R},$ converge strongly to an operator $\mathcal{L}_{\infty}$, as 
$t\rightarrow \infty$, 

\hspace{0.6cm}and there exists a finite constant $C$ such that, uniformly in the region $\Lambda\subset \mathbb{E}^{3}$

\hspace{0.6cm}$S$ is confined to,
\begin{equation}\label{Liouville bound}
\int^{\infty} \Vert \big[\mathcal{L}_{\tau} - \mathcal{L}_{\infty} \big]\big(\mathcal{L}_{\infty}+i\big)^{-1} \Vert \, d\tau \leq C\,.
\end{equation}
}
\item[\quad(ii)]{In the thermodynamic limit ($\Lambda \nearrow \mathbb{E}^{3}$), the spectrum of $\mathcal{L}(t)$ is 
absolutely

\hspace{0.6cm} continuous, except for a simple eigenvalue 0, for all times $t$. }
\end{itemize}

\textit{Remark:} When attempting to prove property (ii) for concrete models one usually 
has to assume that the Hamiltonians $H_C(t)$ satisfy a ``Fermi Golden Rule''-type condition, 
for all times $t\in \mathbb{R}$,\footnote{I am grateful to \textit{Israel Michael Sigal} for having taught 
me how to use Fermi-Golden-Rule conditions in the analysis of quantum-mechanical resonances and spectral 
problems.} and that their norms, $\Vert H_{C}(t) \Vert$, are sufficiently small, uniformly in $t$. 
Furthermore, for the time being, one has to assume that the reservoir $R$ has good dispersive properties; e.g., that it is described by an ideal quantum gas. 

The following result provides a first example of \textit{irreversible behavior}.

\vspace{0.2cm}
\noindent{\textbf{Return to Equilibrium.}}\\
\textit{We suppose that the thermodynamic limit of the system $S$ has been constructed, and that
 properties (i) and (ii) stated above hold.
Let the initial state, $\omega$, of $S$ at some time $t_0$ be given by
$$\omega(A)= \omega^{\beta}_{t_0}\big(M^{*}\, A\, M\big)/\omega^{\beta}_{t_0}(M^{*}M)\,, \,\,\quad A\in \mathcal{A}\otimes B(\mathcal{H}_S)\,,$$
where $\mathcal{A}$ is the algebra defined in \eqref{quasi-loc alg}, $\omega^{\beta}_{t_0}$ is the instantaneous equilibrium state at some inverse temperature $\beta<\infty$
(see Eq.~\eqref{states}), and $M$ is an arbitrary quasi-local operator. Let $\omega_t$ be the true state of $S$ 
at time $t$, with initial condition $\omega_{t_0}=\omega$.\\
Then}
\begin{equation}
\underset{t\rightarrow \infty}{\text{lim}}\, \omega_{t}(A) = \omega^{\beta}_{\infty}(A)\,, \quad \forall\,\, 
\textit{ quasi-local operators }\,\,A\,,
\end{equation}
\textit{where $\omega^{\beta}_{\infty}$ is the equilibrium state at inverse temperature $\beta$ corresponding 
to the limiting Hamiltonian $H_{\infty}$.}

\vspace{0.2cm}\textit{Remarks:} (1) Mathematical study of this phenomenon was initiated in \cite{L-Rob}.

(2) Given properties (i) and (ii), above, Return to Equilibrium is a simple consequence of the KMS condition 
\cite{Br-Rob} (exercise). The hard problem is to verify property (ii) for concrete models.

(3) `Return to Equilibrium' can be extended in various ways. For example, one can prove a result on \textit{thermal 
ionization of atoms} immersed in a heat bath (see \cite{FMS}); and, assuming that $H_C(t)$ is periodic in time $t$ 
with period $\tau$, one can derive a result concerning the approach to time-$\tau$-periodic states after having taken the
thermodynamic limit (see \cite{Abou-F}). 

(4) An irreversible phenomenon more subtle, mathematically, than `Return to Equilibrium' is the phenomenon of 
\textit{Relaxation} of isolated systems (such as atoms or molecules) coupled to the quantized electromagnetic 
field \textit{to their Groundstates.} In some models of atoms or molecules prepared in states with an energy below the 
ionization threshold, various rigorous results on \textit{Rayleigh scattering} of light, the \textit{Bohr frequency 
condition}, and \textit{relaxation to the groundstate} have been established; see 
\cite{FGSch, BFP, DK, FauS}. In these models, the atomic nuclei are assumed to be static, the electrons 
have non-relativistic kinematics, pair-creation is neglected, and the interactions of 
electrons with the quantized electromagnetic field are cut off at high energies 
(i.e., are regularized in the ultraviolet). The phenomenon of relaxation to the groundstate 
is a consequence of \textit{spontaneous emission} of photons by electrons in excited states of the atom (as 
originally predicted by Einstein in 1916). Although this process has been understood at some heuristic level 
for a long time, the job to understand it mathematically rigorously has turned out to be quite involved.
\footnote{An important role in this analysis is played by the fact that, in Rayleigh scattering, a genuine 
infrared problem does \textit{not} arise -- the number of photons created in such processes has a finite 
expectation value (see \cite{DK}), a conjecture originally proposed by the author.}

\vspace{0.2cm} In thermodynamics, \textit{quasi-static} processes, and in particular quasi-static 
\textit{isother- mal} processes (which are
\textit{reversible}), play an important role. Such processes can be described very nicely within 
quantum statistical mechanics. In \cite{Abou-F-2} the following variant of the \textit{adiabatic theorem of quantum mechanics} 
has been proven.

\vspace{0.2cm}
\noindent \textbf{Isothermal Theorem.}\\
\textit{Consider a system $S$ in the thermodynamic limit ($\Lambda \nearrow \mathbb{E}^{3}$),
as above. We choose the Hamiltonian of $C$ to be given by}
$$H_C(t)\equiv H_C^{(\tau)}(t):= K_C(\tau^{-1}\cdot t)\,, \quad \tau > 0\,,$$ 
\textit{where  $K_C(t)$ is $\tau$-independent. We assume that the corresponding Liouvillian, 
$\mathcal{L}_t \equiv \mathcal{L}^{(\tau)}_{t}:= \mathcal{K}_{t/\tau}$, of $S$ (see \eqref{Liouville})
satisfies a number of conditions specified in \cite{Abou-F-2}, including self-adjointness, differentiability 
properties in time and spectral properties (in particular, that 0 is a simple eigenvalue of 
$\mathcal{K}_t$, for all times $t$). Let us suppose that the true state of $S$ at some initial time $t_0$ 
is given by an instantaneous equilibrium state at some inverse temperature $\beta < \infty$, i.e., }
$\omega=\omega_{t_0} = \omega^{\beta}_{t_0}$.\\
\textit{Then, for $t$ in an arbitrary finite interval of length $\propto \tau$ of the time axis $\mathbb{R}$, 
the true state $\omega_t$ of $S$ follows the instantaneous equilibrium state 
$\omega^{\beta}_t$ corresponding to the Hamiltonian $H^{(\tau)}(t)=H_C^{(\tau)}(t) + H_R$, up to an error functional whose 
norm converges to 0, as the adiabatic time scale $\tau$ of the process tends to $\infty$.} \hspace{3.7cm} $\square$

\vspace{0.2cm}This result (whose precise statement and proof can be found in \cite{Abou-F-2}) has an interesting application in
thermodynamics: Let $t_j= \tau\cdot s_j, j=1,2,$ with $s_1 < s_2,$ and let $H_C(t):= H_C^{(\tau)}(t) = K_C(\tau^{-1}t)$. Then
\begin{align*}
\Delta_{t_1}^{t_2}U_{C} + \Delta_{t_1}^{t_2}Q\, =& \,\,\Delta_{t_1}^{t_2}W 
\overset{\eqref{first law}}{=}\int_{t_1}^{t_2} \omega_{t}\big(\dot{H}_{C}^{(\tau)}(t)\big)\, dt \nonumber\\
=&\int_{t_1}^{t_2} \omega_{t}\big(\frac{d}{dt} [K_{C}(\tau^{-1}t)]\big)\,dt
\end{align*}
Assuming that the initial state is an instantaneous equilibrium state, $\omega_{t_1} = \omega_{t_1}^{\beta}$, and
applying the Isothermal Theorem, we conclude that
\begin{align}
\hspace{1cm} \Delta_{t_1}^{t_2}W=&\int_{t_1}^{t_2} \omega^{\beta}_{t}\big(\frac{d}{dt} [K_{C}(\tau^{-1} t)]\big)\,dt 
+ \mathfrak{o}(1) \nonumber\\
=& \Delta_{t_1}^{t_2} F_{C} + \mathfrak{o}(1)\,, \label{free energy}
\end{align}
\[\text{i.e.,}\,\,\, \Delta W = \Delta F_C + \mathfrak{o}(1), \quad\text{where }\,\,\, 
F_{C}(t):= \beta^{-1} \big[\text{ln}Z_{t}(\beta) - \text{ln  Tr}\big(\text{exp}[-\beta H_R]\big)\big]\]
is the \textit{free energy} of the small system $C$ (coupled to the thermal reservoir).
In words: In a \textit{quasi-static isothermal (hence reversible) 
process}, the amount of work done on the subsystem $C$ equals the change of its free energy.

We conclude this subsection with some remarks about \textit{entropy.} When a ``small'' system $C$ is in contact with a
macroscopically large environment $R$ the \textit{entropy} of $C$ should be defined to be the negative \textit{relative
entropy} of the true state, $\Omega_t$, of the total system $S=C\vee R$, given a reference state, $\Omega^{\text{ref}}$, 
with the property that $C$ and $R$ are decoupled from one another. If $R$ is a thermal reservoir then it is assumed to 
be in thermal equilibrium at some inverse temperature $\beta$ before being coupled to $C$ (\textit{Zero$^{th}$ Law}).
Since we have chosen the state space of $C$ to be finite-dimensional, we may choose the normalized trace, tr,
as the reference state of $C$ and set
$$\Omega^{\text{ref}}:= Z(\beta)^{-1} \text{exp}[-\beta H_R] \otimes \mathbf{1}\vert_{\mathcal{H}_C}\,, \,\,\, \text{with }\,\,
Z(\beta) = \text{Tr}\big(\text{exp}[-\beta H_R]\otimes \mathbf{1}\vert_{\mathcal{H}_C}\big)\,. $$
We define the entropy of $C$ at time $t$ to be given by
\begin{equation}\label{entropy of C}
S_C(t):= -k_B \underset{\Lambda \nearrow \mathbb{E}^{3}}{\text{lim}} \text{Tr}\Big(\Omega_t \big[\text{ln} \Omega_t 
- \text{ln} \Omega^{\text{ref}}\big]\Big) \overset{\eqref{rel entropy}}{\leq} 0\,.
\end{equation}
We note that $\text{Tr}\big(\Omega_t \text{ln}\Omega_t \big)$ and $\text{Tr}\big( \Omega_t \cdot Z(\beta)\mathbf{1}\big)$ 
are \textit{independent} of time $t$. Hence
\begin{equation}\label{entropy increase}
\dot{S}_{C}(t)= -\frac{1}{T} \frac{d}{dt}\omega_{t}\big(H_R\big) =- \frac{1}{T} \dot{Q}(t)\,, \quad \text{or}\quad \Delta S_C
= -\frac{1}{T} \Delta Q= \frac{1}{T}\Delta Q^{\swarrow}\,,
\end{equation}
as expected, where $T=(k_B \beta)^{-1}$ is the absolute temperature.

It is instructive to specialize this equation to entropy changes observed in quasi-static isothermal processes.
In such processes, 
$$\omega_t=\omega^{\beta}_t\,,$$
up to small error terms, where $\omega^{\beta}_t$ is the (thermodynamic limit of the) instantaneous equilibrium state introduced in Eq.~\eqref{states}.
Specializing the definition of $S_C$ in \eqref{entropy of C} to this choice of states, one finds that
$$T \dot{S}_C(t) = \frac{d}{dt} \big[\omega^{\beta}_t \big(H_C (t)\big) \big]- \omega^{\beta}_t\big(\dot{H}_C (t)\big)\,,$$
which, with \eqref{first law} and after intergrating over time, yields 
\begin{align}
 \Delta U_C = T\Delta S_C + \Delta W
 \overset{\eqref{entropy increase}}{\quad=}&\,\, \Delta Q^{\swarrow} + \Delta W\,, \quad\text{or, \,with \eqref{free energy}\,,}\nonumber\\
 \Delta F_C =&\,\, \Delta U_C - T \Delta S_C\,.\label{sec law}
\end{align}

To conclude this subsection, we emphasize that, from the point of view of mathe- matical analysis, the hard
problems are (1) to analyze the spectrum of the Liouvillians $\big(\mathcal{L}_t\big)_{t \in \mathbb{R}}$ in the 
thermodynamic limit of the reservoirs and to establish property (ii) after Eq.~\eqref{Liouville bound}, and 
(2) to prove the \textit{Isothermal Theorem}.

\vspace{0.2cm}
\noindent {\bf{\hspace{0.13cm}II. The Second Law of thermodynamics in the formulation due to Clausius}}

We consider a physical system, $S$, consisting of two 
macroscopic thermal reservoirs, $R_1$ and $R_2$, at temperatures $T_1 \,>\,T_2$, respectively, 
coupled to one another by a ``thermal contact'', $C$, i.e., a small system separated from $R_1$ 
and $R_2$ by diathermal walls:
$$S = R_1 \vee C \vee R_2\,.$$
We describe $S$ quantum-mechanically, supposing, for example, that the reservoirs are filled with ideal
quantum gases confined to some large disjoint regions, $\Lambda_1$ and $\Lambda_2$, respectively, 
in physical space (which, later on, will approach infinitely extended half spaces in $\mathbb{E}^{3}$). 
The state spaces of the reservoirs are denoted by $\mathcal{H}_1$ and $\mathcal{H}_2$, 
the state space of $C$ by $\mathcal{H}_C$, which, for simplicity, we assume to be finite-dimensional.
The Hamilton operators describing the time evolution of $R_1$ and $R_2$ are denoted by $H_1$ and $H_2$, 
respectively; the Hamiltonian of $C$, which \textit{includes interaction terms} between $C$ and $R_1 \vee R_2$, 
by $H_{C}$. As in {\textbf{I}}, above, $H_C$ is assumed to be a bounded quasi-local operator acting on 
$\mathcal{H}_S := \mathcal{H}_1 \otimes \mathcal{H}_C \otimes \mathcal{H}_2$.

The total Hamiltonian of $S$ is given by the operator
$$H := H_1 + H_2 + H_{C}$$
on $\mathcal{H}_S$ (with $H_1$ identified with 
$H_1\vert_{\mathcal{H}_1} \otimes \mathbf{1}\vert_{\mathcal{H}_C} \otimes \mathbf{1}\vert_{\mathcal{H}_2}$, 
and similarly for $H_2$).

The state of $S$ at a time $t\in \mathbb{R}$ is given by a \textit{density matrix}, $\Omega_t$, see {\bf{I}}.
Before $C$ is ``opened'', this state is assumed to be given by a tensor product of \textit{Gibbs states} 
(as suggested by the \textit{Zero$^{th}$ \,Law} of thermodynamics):
\begin{equation}\label{product state}
\Omega^{\text{ref}}:= Z^{-1} exp(-\beta_{1}H_{1})\otimes {\bf{1}}_{C} \otimes exp(-\beta_{2}H_{2}),
\text{   }\,\, \beta_{i}:=1/k_{B}T_{i}\,.
\end{equation}
\begin{center}
\includegraphics[width=5cm]{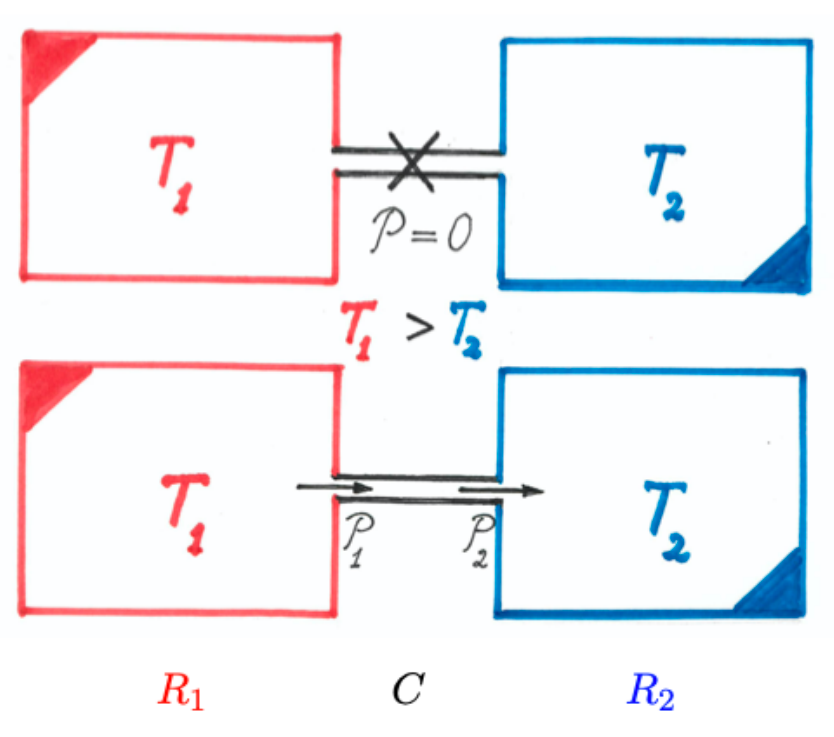}\\
{\small{\textit{Figure 1:} Reservoirs coupled by a thermal contact}}
\end{center}
The {heat power}, $\mathcal{P}_{i}$, absorbed by $R_i, \,i=1,2,$\, is defined by
\begin{equation}\label{power}
\mathcal{P}_{i}(t):= \frac{d}{dt}\text{Tr}(\Omega_{t}H_{i}) = -i\text{Tr}\big(\Omega_{t}[H_{i},H_{C}]\big)\,.
\end{equation}
(Since the operators $[H_i, H_C]$ are quasi-local, the thermodynamic limit of right side of \eqref{power} may exist.)  
If $R_1$ and $R_2$ have ``good dispersive properties'', e.g., are filled with an 
ideal quantum gas or with black-body radiation, one proves that, in the thermodynamic limit,
$$\omega_t \underset{t\rightarrow \infty}{\rightarrow} \omega_{\infty}\,,$$
where $\omega_{\infty}$ is a so-called \textit{non-equilibrium stationary state} (NESS). This is among the
somewhat hard results proven in this context; see \cite{FMUe, JP-2} and followers.\footnote{An early analysis of such states
appeared in \cite{Lebowitz-2}. A mathematically precise result of this type was established 
in the diploma thesis of S. Dirren, written under the supervision of G.~M.~Graf and the author 
(ETH Zurich, winter 1998/99), following ideas and methods in \cite{L-Rob}.}

As in Sect.~1.2 and \textbf{I}, we consider the relative entropy 
\begin{equation*}
S(\Omega_t \Vert  \Omega^{\text{ref}})=\text{Tr}\big(\Omega_{t}[\text{ln}\Omega_{t}-\text{ln}\Omega^{\text{ref}}]\big)\,,
\end{equation*}
with $\Omega^{\text{ref}}$ as in \eqref{product state}. It is easy to see that its time derivative is given by
\begin{equation}\label{entropy id}
\dot{S}(\Omega_{t} \Vert  \Omega^{\text{ref}})= \beta_{1}\mathcal{P}_{1}(t) + \beta_{2} \mathcal{P}_{2}(t)\,.
\end{equation}

\noindent Thus, if $\omega_{t} \rightarrow \omega_{\infty}$\,, as $t\rightarrow \infty$, with $\omega_{\infty}$ \textit{time-translation-invariant}, then the following statements hold in the thermodynamic limit:
\begin{itemize}
\item[1.]{$\dot{S}(\Omega_{t} \Vert  \Omega^{\text{ref}})$ has a limit, denoted by $\sigma_{\infty}$, as $t \rightarrow \infty$, 
and it follows from the non-negativity of relative entropy, see \eqref{positivity}, that 
\begin{equation}\label{entropy prod}
\,\sigma_{\infty} \geq 0 \qquad\text{(positivity of entropy production)}
\end{equation}
and}
\item[2.]{$\mathcal{P}_{1}(t)$ has a limit, denoted by $-\mathcal{P}_{\infty}$, as $t \rightarrow \infty$, and
\begin{equation}\label{energy balance}
\mathcal{P}_1(t) + \mathcal{P}_2(t) \rightarrow 0\,, \,\,\text{    as   }\,\, t \rightarrow \infty.
\end{equation}
}
\end{itemize}
From Eqs.~\eqref{entropy id}, \eqref{entropy prod} and \eqref{energy balance} we derive the \textit{Second Law} 
of thermodynamics in the formulation of Clausius:
\begin{equation}\label{heat flow}
\mathcal{P}_{\infty} \underbrace{(\frac{1}{T_2}-\frac{1}{T_1})}_{>0} \geq 0\quad \Rightarrow \quad \mathcal{P}_{\infty} \geq 0,
\end{equation}
i.e., at large times when a stationary state has been reached, \textit{heat flows from the warmer reservoir, $R_1$, 
to the colder one, $R_2$}. This is the Second Law of thermodynamics in the formulation of Clausius.

So far, we have worked in the \textit{canonical ensemble}: The subsystem $C$ is connected to thermal reservoirs only
through \textit{diathermal walls}; matter is not exchanged between $C$ and the reservoirs. Actually, it is easy to extend
our analysis and the results reviewed so far to the \textit{grand-canonical ensemble}, with subsystems connected to reservoirs 
through channels transmitting not only energy but also matter (atoms or molecules); see, e.g., \cite{FMUe}. 

\vspace{0.2cm}\noindent \textit{Further results:} 
\begin{itemize}
\item{For certain simple models one can show that, at large times, the entropy production $\sigma_{\infty}$ 
is strictly positive, and }
\item{\textit{Ohm's Law} holds; see \cite{FMUe}.}
\item{The \textit{Onsager relations} hold; see \cite{Hepp, JOP}.}
\end{itemize}

\vspace{0.2cm}
\noindent \textbf{III. The Second Law in the formulation of Carnot}

We may replace the thermal contact $C$ between two thermal reservoirs $R_1$ and $R_2$ by a \textit{heat engine}, 
e.g., a ``locomotive'', $E$, that extracts energy from a heater $R_1$, releases part of it into a ``cooler'' $R_2$,
with $T_1 > T_2$, and performs work. Thus, the system we consider is
 $S=R_1 \vee E \vee R_2$.
 
The engine $E$ is driven \textit{periodically in time}, with some period $\tau>0$. This means that the 
 Hamiltonian, $H_{E}(t)$, depends periodically on time $t$, with period $\tau$. Assuming that $R_1$ and $R_2$ have 
 good dispersive properties, e.g., are filled with ideal quantum gases or black-body radiation, and developing a 
 \textit{Floquet theory} for Liouvillians (see \cite{Abou-F}), one proves that, in the thermodynamic limit, 
 the true state, $\omega_{t}$, of $S$ approaches a \textit{time-periodic state}, $\omega^{\text{asy}}_t$, 
 with the \textit{same} period, $\tau$, as the Hamiltonian of $E$. 

Using this fact, we are able to derive \textit{Carnot}'s bound on the degree of efficiency, $\eta$, of heat engines.
The existence of time-periodic asymptotic states $\omega^{\text{asy}}_t$ and the non-negativity of relative entropies,
i.e.,  $S(\Omega_t \Vert \Omega^{\text{ref}}_t) \geq 0$,  with $\Omega^{\text{ref}}$ as in Eq.~\eqref{product state} (with
$C$ replaced by $E$) imply that, asymptotically as time $t\rightarrow \infty$, the
\textit{entropy production per cycle} (of length $\tau$), $\Delta S_{\infty}\equiv \Delta S_{\infty}(\tau)$, is \textit{non-negative}.
Let $\Delta Q_{1}^{\swarrow}$ be the amount of heat energy released per cycle by the heater $R_1$ into $E$, 
and let $\Delta Q_2^{\nearrow}$ be the heat energy lost per cycle by $E$ into the cooler $R_2$, 
after the state of $S$ has approached $\omega_t^{\text{asy}}$. Using the identity
$$0\leq \Delta S_{\infty} = - \frac{\Delta Q_{1}^{\swarrow}}{T_1} + \frac{\Delta Q_2^{\nearrow}}{T_2}$$
(see \eqref{entropy id}) we conclude that \textit{Carnot's bound} on the \textit{degree of efficiency} holds.

\vspace{0.2cm}
\noindent \textbf{Theorem.} (Carnot's bound) \\
\textit{The degree of efficiency, $\eta$, of the heat engine $E$ satisifies the
familiar bound}
\begin{align}\label{Carnot}
\eta:=  \frac{\Delta W}{\Delta Q_1^{\swarrow}} =& \frac{\Delta Q_1^{\swarrow}-\Delta Q_2^{\nearrow}}{\Delta Q_1^{\swarrow}} \leq \frac{T_1 - T_2}{T_1}\equiv\,\, \eta_{\text{ Carnot}}\,, \,\,\text{and}\nonumber\\
 \eta\,= &\,\,\eta_{\text{ Carnot}} \,\,\text{ iff }\,\, \Delta S_{\infty}(\tau)=0\,.
\end{align}
This is \textit{Carnot}'s formulation of the \textit{$2^{nd}$ Law} of thermodynamics. (Note that, since, asymptotically, 
the state of $S$ approaches one that is periodic in time, the change of the internal energy of $E$ per cycle, 
$\Delta U_E$, very nearly vanishes at very large times. This explains why $\Delta U_E$ does not appear 
in the formula for $\eta$.)

\vspace{0.15cm}
\noindent \textit{Remark:} Defining the \textit{entropy of a small system}, $E$, coupled to an unobserved environment, $R$ 
(such as some thermal reservoirs), as (-1\,$ \times$) the \textit{relative entropy} of the state of the total system, $S=E \vee R$, with respect 
to a reference state in which $E$ is decoupled from $R$, we find that, in stationary or time-periodic states, which are approached
asymptotically, the \textit{entropy production} is always \textit{non-negative} (and strictly positive in certain simple 
model systems). This fact implies that in irreversible adiabatic processes of the small system $E$ (defined appropriately) 
its entropy can \textit{never decrease} (actually increases in certain simple model systems) -- as expected on the basis of 
standard formulations of the \textit{Second Law}.

\vspace{0.2cm}
\noindent {\textbf{IV. The Zero$^{th}$ Law of thermodynamics}}

We conclude this section with a comment on the \textbf{0$^{th}$ Law} of thermodynamics (which is probably the 
deepest law of thermodynamics). It can be formulated 
as follows: Assume that the particle- and energy \mbox{densities} of the initial state of a macroscopic system, $R$, 
are uniformly bounded in space, and that the dynamics of $R$ is translation-invariant and time-independent. Then, 
asymptotically as time $t$ tends to $\infty$ and as the thermodynamic limit is approached, the state of $R$ 
approaches a state \textit{locally indistinguishable} from an \textit{equilibrium state} whose particle- and energy densities 
are given by the spatial averages of the particle- and energy densities of the initial state.

A fundamental problem is to find the right hypotheses on the properties of $R$ enabling one to establish the validity 
of the \textbf{0$^{th}$ Law} in a form involving realistic time scales, and hence to show that 
heat baths or thermal reservoirs really exist. (This can also be phrased as the problem of ``eth'' vs. many-body localization).
The 0$^{th}$ Law represents a key problem encountered in trying to derive the laws of thermodynamics from statistical mechanics
that is not understood very well, yet. Preliminary results can be found in \cite{zero law}.

\section{Quantum Brownian Motion}\label{QBM}
In this section we describe results, formulated and established within quantum mechanics, concerning an example 
of a physical system exhibiting an \textit{arrow of time} (in the sense specified at the beginning of this paper), namely the
example of \textit{Brownian motion} of a particle coupled to a thermal reservoir.

As is well known, the theoretical study of Brownian motion was initiated by \textit{Einstein} and \textit{Smoluchowski} 
in 1905, and their work played a significant role in finally establishing the atomistic nature of matter in experiments 
by \textit{Perrin} and others. But it took a century before one succeeded in establishing the diffusive nature of motion 
of a quantum-mechanical particle interacting with a translation-invariant, infinitely extended quantum-mechanical 
thermal reservoir (disregarding from results on simple exactly solved model systems; see, e.g., \cite{Legget}, 
and references given there). Here we describe some fairly recent results on diffusive motion that 
have appeared in \cite{DeR-F, DeR-K}.

We consider a quantum-mechanical tracer particle hopping on a lattice $\mathbb{Z}^{3}$ that has two internal states,
a groundstate and an excited state. Its Hilbert space of pure state vectors is given by
$\mathcal{H}_P:=\ell^{2}(\mathbb{Z}^{3})\otimes\mathbb{C}^{2}$, and its Hamiltonian is given by
\begin{equation}\label{particle ham}
H_{P}:= -\frac{\Delta_X}{2M}\otimes{\bf{1}}+{\bf{1}}\otimes \sigma^{z}, 
\end{equation}
where $M$ is the mass of the tracer particle, $\Delta_X$ is the discrete Laplacian, $X$ is the position operator 
of the particle, with $\text{spec}X = \mathbb{Z}^{3}$, and $\sigma^{z}=\begin{pmatrix} 1 & 0\\ 0& -1 \end{pmatrix}$\,; we
work in units where $\hbar =1$.

The tracer particle is immersed in a \textit{Bose gas}.\footnote{See \cite{LSSY} for comprehensive information
on the mathematical theory of Bose gases -- one of Elliott Lieb's many interests!} The atoms constituting  the 
Bose gas are \textit{free, non-relativistic particles} with Bose-Einstein statistics and a mass 
$m =\frac{1}{2}$ ($m\ll M$) propagating freely in $\mathbb{E}^{3}$. The dynamics of the 
Bose gas is described by the usual Hamiltonian, denoted by $H_{BG}$, which is self-adjoint 
and non-negative on the bosonic Fock space $\mathfrak{F}$. 

The Hamiltonian of the coupled system, $S=$ \textit{tracer particle $\vee$ Bose gas}, is given by
\begin{equation}\tag{14}
H:= H_P + H_{BG} + \nu \int_{\mathbb{R}^{3}} dx \text{   }W(X-x)\lbrace b^{*}(x)+b(x)\rbrace ,
\end{equation}
acting on the Hilbert space $\mathcal{H}_P \otimes \mathfrak{F}$, where $W$ describes interactions of the tracer 
particle with the atoms of the Bose gas; it is given by some $2\times2$ matrix-valued pair potential that is chosen 
in such a way that it can cause transitions between the groundstate and the excited state of the particle 
(Fermi's Golden Rule!); the coupling constant $\nu$ is given by $\nu:=\sqrt {\rho_{0}/2}$, with $\rho_0$ a
dimensionless quantity proportional to the \textit{density} of the Bose gas, and $b^{*}(x)$ and $b(x)$ 
are creation- and annihilation operators of \textit{phonons} (= quanta of \textit{sound waves}) in the Bose 
gas satisfying the usual canonical commutation relations.

\vspace{0.2cm}
\noindent We distinguish two regimes:

(A) {$\nu$ small}, $M=\nu^{-2}M_{0}$, where $M_{0}$ is a constant\,\, (kinetic regime); and

(B) {$\nu$ large}, with $M=\nu^{2}M_{0}$, \,\,$\nu^{-2} \leftrightarrow \hbar$\,\, (mean-field regime).

\vspace{0.2cm}
We first study regime (A), assuming that the Bose gas is in \textit{thermal equilibrium} at some finite positive temperature 
$T=(k_{B}\beta)^{-1}$. The problem of interest to us is to understand the nature of the motion of the tracer particle.
For this purpose we will study the \textit{diffusion constant}, $D$, which characterizes its large-time behavior. 

We prepare the system in an initial state of the form $\rho_T:=\sigma_0 \otimes \omega^{\beta}$ ($\beta=1/k_B T$),
where $\sigma_0$ is given by a density matrix, $\Sigma_0$, on $\mathcal{H}_P$ localized near $X=0$ (i.e., 
$\text{Tr}\big(\Sigma_0 X^{2}\big)$ is finite). We study this system in the \textit{Heisenberg picture}: 
The state $\rho_T$ is taken to be time-independent, but the operators representing physical quantities 
of the system evolve in time according to Heisenberg's equations. Thus,
$X(t)$ is the operator representing the position of the tracer particle at time $t$.
By $\langle (\cdot) \rangle_{T}$ we denote expectations with respect to the state $\rho_T$.

One attempts to prove and, after much hard work, succeeds in proving \cite{DeR-F, DeR-K} that
\begin{equation}\label{diffusion const}
\langle [X(t)-X(0)]^{2}\rangle_{T} \sim D\cdot t, \quad\text{    as  }\,\, t \rightarrow \infty,
\end{equation}
for a diffusion constant $D$ proportional to $\nu^{2}$, as can be guessed on the basis of simple dimensional
analysis:
$$D \approx \big[(\overline{v}_{\nu} \times \overline{t}_{\nu})^{2} /\overline{t}_{\nu}\big] \, \propto \, \nu^{2}< \infty\,,$$
where $\overline{v}_{\nu} \propto \nu^{2}$ is the average speed of the particle (recall that its mass is 
proportional to $\nu^{-2}$), and $\overline{t}_{\nu} \propto \nu^{-2}$ is the average time elapsing 
between two subsequent collisions of the particle with an atom in the Bose gas (recall that the strength of
interaction between the particle and sound waves in the Bose gas is $\mathcal{O}(\nu)$).

\vspace{0.2cm}\noindent \textit{Rough idea of the proof of} \eqref{diffusion const}: Evidently
\begin{equation}\label{int rep}
\langle [X(t)-X(0)]^{2}\rangle_{T} = \int_{0}^{t} d\tau \int_{0}^{t} d\sigma\, \langle \dot{X}(\tau)\cdot 
\dot{X}(\sigma) \rangle_{T}\,.
\end{equation}
Assuming that $\langle \dot{X}(\tau)\cdot \dot{X}(\sigma) \rangle_{T}$ decays integrably fast in $\vert \tau - \sigma \vert$,
the right side of \eqref{int rep} grows \textit{linearly} in $t$, as $t \rightarrow \infty$, which implies that
$D < \infty$.

 Integrable decay, with a decay time of $\mathcal{O}(\nu^{2})$, is due to fast de-correlation (in time) of the 
 direction of motion of the particle as a result of collisions with atoms in the Bose gas. This can be established,
 mathematically, with the help of a rather intricate ``cluster expansion in time,'' which yields precise information
on quantities such as $\langle \dot{X}(\tau)\cdot \dot{X}(\sigma) \rangle_{T}$. 

Some key ideas underlying this expansion are as follows: 
Let $\mathcal{Z}^{\nu}_t$ denote the completely positive map acting on the algebra, $B(\mathcal{H}_P)$, 
of bounded operators on $\mathcal{H}_P$ that takes an operator $A\equiv A(0)$ representing a physical 
quantity, $\widehat{A}$, characteristic of the tracer particle at time $0$ to the operator $A(t)$ representing 
$\widehat{A}$ at time $t$; assuming that the degrees of freedom of the heat bath (the Bose gas) have been 
``traced out.'' If the dynamics of the particle were \textit{ballistic} then 
 $$A(t)=\mathcal{Z}^{\nu}_t(A) \approx \text{exp}[itH_P^{\text{ren}}]\,A\,\text{exp}[-itH_P^{\text{ren}}], \,\,\text{ and }\,\,
 \langle [X(t)-X(0) ]^{2}\rangle_T \propto \nu^{4}\cdot t^{2},$$
for an effective (renormalized) Hamiltonian $H_P^{\text{ren}} \propto -\nu^{2} \Delta_X$. Instead, thanks to the interactions 
of the tracer particle with the atoms in the Bose gas, which are in a thermal state at temperature $T$, 
one expects that
 \begin{equation}\label{Lindblad}
 A(t) = \mathcal{Z}^{\nu}_t(A) \approx \text{exp}\big[t\big(i\,ad_{\sigma^{z}}+\nu^{2}\mathcal{M}_{T}\big)\big](A), 
 \quad A=A(0)\in B(\mathcal{H}_P),
 \end{equation}
 where $\mathcal{M}_{T}$ is the generator of a semi-group of completely positive maps on $B(\mathcal{H}_P)$
 to be chosen selfconsistently; (it is related to a linear Boltzmann equation for the Wigner distribution of the state
 $\sigma_t$ of the tracer particle in the Schr\"odinger picture). If the maps
 $\text{exp}\big[t\big(i\,ad_{\sigma^{z}}+\nu^{2}\mathcal{M}_{T}\big)\big], t\geq 0,$ gave rise to the exact time evolution 
 of operators related to the tracer particle then Eq. \eqref{diffusion const} would hold, for some positive constant 
 $D=\mathcal{O}(\nu^{2})$. The idea is then to expand the effective dynamics, $\mathcal{Z}^{\nu}_t(\cdot)$, 
 of the tracer particle around the right side of \eqref{Lindblad}. Such an expansion has a chance to converge 
 \textit{uniformly} in time $t$, whereas an expansion of the effective dynamics around the free (ballistic) dynamics 
 of the tracer particle  \textit{cannot} converge uniformly in $t$, because the latter has the wrong behavior for
 very large times. 
  
The reason why we assume that the tracer particle has an internal degree of freedom is that, in this study, 
we would like to profit from the ultraviolet regularization of the problem provided by the lattice. We would therefore 
not like to have to re-scale space and time (in such a way that the continuum limit is approached). 
An internal degree of freedom makes it easy to satisfy momentum- \textit{and} energy 
conservation in collisions of the tracer particle with atoms in the Bose gas; 
(see \cite{DeR-F} for a more detailed explanation).
  
 We note that, in this quantum-mechanical study of diffusive motion, 
 $$\overline{v}_{\nu} \sim \langle \Vert \dot{X}(t) \Vert \rangle_T = \mathcal{O}(\nu^{2}),$$
 i.e., the expected speed of the tracer particle is finite (and small, for small $\nu$).\footnote{Of course, for a particle hopping
 on the lattice $\mathbb{Z}^{3}$ to have a \textit{finite speed} is not surprizing. The \textit{smallness} of $D$ and 
 $\overline{v}_{\nu}$, as $\nu\rightarrow 0,$ is due to the large mass, $M= \mathcal{O}(\nu^{-2})$, of the particle.}
 This is in marked contrast to ordinary Brownian motion for which $\dot{X}(t)$ does not exist.
 Furthermore, a variant of the \textit{Equipartition Theorem} holds for $\Vert \dot{X}(t)\Vert^{2}$.

These results appear to be the \textit{first and only results} on a derivation of diffusive motion from 
fundamental quantum dynamics in a model that cannot be solved exactly (see \cite{DeR-F}; and \cite{DeR-K} for
stronger results and better methods of proof).

It may be of interest to note that if one couples a particle moving in a \textit{disordered (random) potential} 
to a thermal reservoir then (Anderson) \textit{localization breaks down}, and the particle exhibits diffusive motion, no
matter how weak the coupling of the particle to the thermal reservoir is. Suppose the disorder in the random 
potential is large enough for the particle to be localized before it is coupled to the reservoir, no matter what its 
initial energy is. Then the diffusion 
constant tends to 0, as the strength of the coupling of the particle to the reservoir tends to 0; not because the 
mass of the particle diverges (as in the kinetic regime (A)), but because of Anderson localization in the 
absence of thermal noise. Results like this have been proven in \cite{F-Schenker} for simplified models 
with Markovian thermal reservoirs.

\section{Hamiltonian Friction}\label{Friction}
\textit{``A moving body will come to rest as soon as the force pushing it no longer acts on it in the manner necessary for its propulsion.''} (Aristotle)

In this section we consider an archetypal example of an \textit{irreversible phenomenon}, namely \textit{friction},
in a model of a tracer particle propagating in an ideal Bose gas at zero temperature exhibiting Bose-Einstein 
condensation; (see \cite{Pulvi-1, Pulvi-2, F-GZ-1} for various results on friction in mechanical models). 
This is a problem in \textit{``tribology,''} a branch of physics first explored by Aristotle, Leonardo da Vinci
and Guillaume Amontons. In our model friction arises because the particle emits Cherenkov radiation 
of sound waves into the condensate of the Bose gas.\footnote{Cherenkov radiation is observed 
in nuclear reactors: Electrons moving through water at a speed larger than the speed of 
light in water emit blue light. This causes them to decelerate until their speed has dropped to below 
the speed of light in water.} To be able to come up with mathematically precise results, we study the 
model in the mean-field regime introduced in Sect.~4, regime (B), where the mass, $M$, of the particle 
behaves like $M=\nu^{2}M_{0}, M_0 = \mathcal{O}(1),$ $\nu^{-2} \sim \rho_0^{-1} \leftrightarrow \hbar$, 
with $\nu^{-2} \rightarrow 0$. We consider a particle \textit{without} internal degrees of freedom 
moving in physical space $\mathbb{E}^{3}$, rather than hopping on $\mathbb{Z}^{3}$. 

We suppose that the particle interacts with the atoms of the Bose gas through a \textit{smooth} 
pair potential, $W(X-x)$, of \textit{short range}, where $X\in \mathbb{E}^{3}$ denotes the position 
of the tracer particle, and $x \in \mathbb{E}^{3}$ is the position of one of the atoms in the Bose gas. 
The gas is in a state of high density $\rho_0$ and very low temperature $T\approx 0$, exhibiting a Bose-Einstein 
condensate (i.e., the gas is in its groundstate).

In the mean-field limit described above, the dynamics of this system approaches 
the \textit{classical Hamiltonian dynamics} of a particle, whose state is 
described by its position $X\in \mathbb{E}^{3}$ and momentum $P\in \mathbb{R}^{3}$, 
coupled to a complex-valued (c-number) field, $\beta(x),$ $x \in \mathbb{E}^{3}$, describing sound 
waves in the Bose gas, where  $\beta$ is an element of the Sobolev space $\mathcal{H}^{1}(\mathbb{E}^{3})$. 
The phase space, $\Gamma$, of the system is given by 
 $$\Gamma := \mathbb{R}^{6} \times \mathcal{H}^{1}(\mathbb{E}^{3})\,.$$
Its symplectic structure is encoded in the Poisson brackets
\begin{equation}\label{Poisson brackets}
\lbrace \beta(x), \overline{\beta}(y) \rbrace = i\delta(x-y), \text{    } \lbrace X^{i},P_j \rbrace = -\delta^{i}_{j}\,,
\end{equation}
and
$$\lbrace \beta^{\#}(x), \beta^{\#}(y) \rbrace = \lbrace X^i, X^j \rbrace = \lbrace P_i, P_j \rbrace =0\,, \quad \forall \, x,y,i \text{ and } j.$$
\noindent The Hamilton functional of the system is given by
\begin{align}\label{Hamiltonian}
H_{cl}(X, &P;\beta,\overline{\beta}):= \frac{P^{2}}{2M_0} - F\cdot X + \nonumber\\
+2& \int_{\mathbb{E}^{3}}dx \, W(X-x)\,\text{Re}\beta(x) + \int_{\mathbb{E}^{3}}dx \,
(\nabla \overline{\beta})(x) \cdot (\nabla \beta)(x)\,,
\end{align}
where $F$ is a constant external force gently pushing the particle, and $W$ is a smooth function on 
$\mathbb{E}^{3}$ of short range. The equations of motion of the particle and of the sound-wave field 
$\beta$ of the Bose gas are given by
\begin{equation}\label{eq motion-1}
\dot{X}_t=M_{0}^{-1}P_t, \qquad
\dot{P}_t= F -2 \int_{\mathbb{R}^{3}}dx\text{  }\nabla W(X_{t}-x)\text{ Re}\beta_{t}(x),
\end{equation}
and
\begin{equation}\label{eq motion-2}
i\dot{\beta}_{t}(x)=-\Delta \beta_{t}(x)+ W(X_{t}-x)
\end{equation}

\vspace{0.15cm} \noindent \textit{Remark:} Canonical quantization of the Hamiltonian system \eqref{Poisson brackets} - \eqref{Hamiltonian} 
(with $\hbar^{-1} \sim \nu^{2} \propto \text{mass of particle } \propto$ density of Bose gas) reproduces the 
quantum system that we have started from. One can prove a \textit{Egorov-type theorem} (see \cite{FKS}):

\hspace{0.8cm}\textit{Quantization and time-evolution commute, up to errors of $\mathcal{O}(\nu^{-2})$.}

\vspace{0.1cm} \noindent This means that insights into the dynamics of the classical system reveal features of the quantum-mechanical
dynamics over a finite period of time in a regime of large values of $\nu$.

We first study ``stationary'' solutions of Eqs.~\eqref{eq motion-1} and \eqref{eq motion-2}, setting 
$$\dot{P}_t=0 \,\text{ and }\, \beta_{t}(x)=\gamma_{v}(x-vt-X_{0}), \text{  with  }\, X_{t}=X_{0}+vt,$$
where $v$ is the velocity of the particle (here assumed to be constant), and $\gamma_{v}$ is a function 
determined by plugging the above ansatz into \eqref{eq motion-2}. Eq.~\eqref{eq motion-1} then 
tells us that the external force $F$ must be cancelled by the second term on the right side, 
which describes a \textit{friction force} arising from the particle's emission of 
\textit{Cherenkov radiation} of sound waves into the Bose gas.

\vspace{0.15cm}
\noindent \textbf{Result:} \textit{If $W$ is smooth and of short range then there is a positive constant 
\mbox{$F_{max}<\infty$} such that:}
\begin{itemize}
\item{\textit{If $\vert F \vert < F_{max}$ then there are two stationary solutions of the equations of motion 
describing a particle propagating with a constant velocity $v\Vert F$, with $|v|$ either given by $v_{-}(F)$ 
(a stable solution) or by $v_{+}(F) > v_{-}(F)$ (a ``run-away'' solution), accompanied by a splash 
in the condensate whose shape is given by $\gamma_{v_{\pm}}$. 
The behavior of a particle perpared in an initial state with a speed $v$ close to $v_{+}(F)$ 
may be perceived as paradoxical: When the external force $F$ pushing the particle decreases it 
becomes faster! (See Figure 2 shown below.)}}
\item{\textit{If $\vert F \vert > F_{max}$ then stationary solutions do not exist; \textit{the particle accelerates for ever!}}}
\end{itemize}
\begin{center}
\includegraphics[width= 8.5cm]{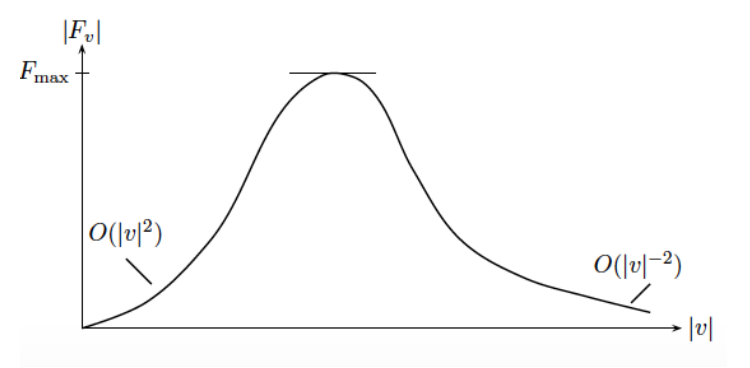}\\
\small{\textit{Figure 2:} External force as a function of paticle speed}
\end{center}
It would be of interest to study the (rate of) approach of a solution of the equations of motion to a 
stationary solution and to test all these predictions in experiments.

\vspace{0.2cm}Next, we study what happens to the particle when the external force $F$ vanishes. As long 
as the speed of the particle is \textit{larger} than the minimal speed of sound, $v_{*}$, in the Bose gas, 
it keeps loosing energy into sound waves, which, thanks to the dispersive properties of the gas, propagate 
to spatial infinity. In an ideal Bose gas, $v_{*}=0$. In this case, the particle will come to rest, as time 
$t$ tends to $\infty,$ as imagined by Aristotle. Here is a theorem proven in \cite{F-GZ-1} (see also \cite{F-G-S}).

\vspace{0.2cm}
\noindent \textbf{Theorem.} (Friction by emission of Cherenkov radiation)\\
 \textit{We consider a particle moving in an ideal Bose gas at zero temperature and interacting 
 with the gas atoms through a two-body potential, $W$, that is smooth and of short range and such that
 $\int_{\mathbb{E}^{3}} W(x)\,d^{3}x \not=0$.
Then there exists a constant $\delta_{*}\approx 0.66$ such that, given an arbitrary $\delta \in (0,\delta_{*})$,  
there exists an $\varepsilon = \varepsilon(\delta)>0$ with the property that,
for initial conditions with 
$$ \Vert (1+ |x|^{2})^{\frac{3}{2}} \beta_{0}(x) \Vert < \varepsilon, \,\, \vert P_{0} \vert < \varepsilon,$$
the solution of the equations of motion \eqref{eq motion-1} and \eqref{eq motion-2} behave as follows:
 \begin{equation}\label{stop}
 \vert P_{t} \vert \leq \mathcal{O} (t^{-\frac{1}{2} - \delta}), \quad \Vert \beta_{t} - \Delta^{-1}W(X_{t}- \cdot)\Vert_{\infty} 
 \rightarrow 0, \quad \text{  as  } t \rightarrow \infty.
 \end{equation}
Choosing $\varepsilon$ so small that $\delta > \frac{1}{2}$, one has that  $X_{t} \rightarrow X_{\infty}$, 
as $t \rightarrow \infty$, with} $\vert X_{\infty} \vert < \infty$, \textit{i.e., the particle comes to rest at a
finite position in physical space, as $t$ tends to $\infty$.}

\vspace{0.3cm}
\noindent \textit{Remark:} An analogous result for the model studied in the kinetic limit can be found 
in \cite{BDeRF}. 

The proof of this theorem is quite technical, and I won't go into any details. But here is an idea. Given a particle trajectory 
$$\underline{X}:=\big\{X_t= X_0+ M_{0}^{-1} \int_{0}^{t} P_s\, ds \, \big| \, 0\leq t < \infty\big\}, $$
where $\underline{P}:= \big\{P_t\,\big|\, 0\leq t \leq \infty\big\}$ is an element of a suitably chosen Banach 
space, we can solve Eq.~\eqref{eq motion-2} for $\beta_{t}(x)=\beta_{t}(x\vert \underline{X})$. We then plug 
the result into the right side of Eq.~\eqref{eq motion-1}, with $F=0$. This yields an integro-differential equation 
with memory of the form
\begin{equation}\label{memory}
\dot{P}_{t}= F(\underline{P}_t,t)+L(\underline{P}_t,t), \quad \text{with }\,\,\, 
\underline{P}_t:= \big\{P_s \,\big|\, 0\leq s\leq t\big\}\,,
\end{equation}
where $F(\underline{P}_t,t)$ is a sum of three terms depending on $\underline{P}_t$ and on $W$, 
one of which depends (\textit{linearly}) on the initial condition, $\beta_0$, of the Bose gas, and where 
$L(\underline{P}_t,t)$ is a sum of two terms depending linearly on $\underline{P}_t$ and quadratically on $W$, 
but \textit{not} on $\beta_0$. In order to show local well-posedness of Eq.~\eqref{memory}, it is convenient to convert it 
into an integral equation for $\underline{P}$.

Let $\mathbb{B}_{\delta}$ be the Banach space of momentum trajectories, $\underline{P}$, with the property
that $\Vert \underline{P} \Vert_{\delta}:= {\text{sup}}_{0\leq t <\infty} (1+t)^{\frac{1}{2}+ \delta} \vert P_t\vert < \infty$, 
where $\delta$ is the parameter appearing in the statement of the theorem. One then shows that, for
$\delta \in (0, \delta_{*})$, $\varepsilon= \varepsilon(\delta)$ can be chosen small enough such that if 
$$ \Vert (1+ |x|^{2})^{\frac{3}{2}} \beta_{0}(x) \Vert < \varepsilon, \,\, \vert P_{0} \vert < \varepsilon,$$
then the integral equation for $\underline{P}$ has a unique solution in $\mathbb{B}_{\delta}$, with 
$\Vert \underline{P} \Vert_{\delta} \leq \mathcal{O}(\varepsilon).$
This is proven by applying a suitable fixed-point theorem on the Banach space $\mathbb{B}_{\delta}$. For all further
details, see \cite{F-GZ-1}.

Results extending those stated in the theorem above have been established for \textit{interacting Bose gases} in the 
Bogoliubov limit, which have the property that the minimal speed of sound, $v_{*},$ is strictly \textit{positive}. 
In this situation a particle prepared in a state with speed below $v_{*}$ moves inertially for ever \cite{F-GZ-2}, 
while a particle prepared in a state with speed larger than $v_{*}$ \textit{decelerates} until its speed approaches 
$v_{*}$, at which point it starts to move inertially \cite{F-GZ-3}. (The proof of this last result is considerably more intricate 
than the proof of the theorem stated above.) Somewhat related, interesting results have been published 
in \cite{Spohn} and references given there.

It would be good to test the theoretical predictions discussed above in experiments.

\section{L'Insoutenable Fl\`eche du Temps dans l'\'{E}volution Quantique}\label{ETH}
\textit{``Alle Naturwissenschaft ist auf die Voraussetzung der vollst\"{a}ndigen kausalen Verkn\"{u}pfung 
jeglichen Geschehens begr\"{u}ndet.''} (Albert Einstein, Zurich physical society, 1910)

\vspace{0.15cm}
\noindent Well, is it?

\vspace{0.1cm}
Analyzing this question within Quantum Mechanics (QM) is the goal of this section, which describes 
some fairly recent insights and ideas described in \cite{BFSch, FP, Fr}. Quite to my surpise, these ideas 
appear to give rise to somewhat violent controversies, which is why I have chosen a somewhat 
shimmering title for this section.\footnote{It is inspired by a novel of \textit{Milan Kundera}, 
entitled: \textit{L'insoutenable l\'eg\`ert\'e de l'\^etre}.}

In Quantum Mechanics, the basic difference between \textit{Past} and \textit{Future} is mirrored in the 
fundamental dichotomy\footnote{According to Edmund Husserl, \textit{actuality} means existence in space and time, 
as opposed to possibility or \textit{potentiality}, which refers to the capacity, power, ability, or chance for something 
to happen or to occur.}
\begin{center}
\textbf{Past} = \textit{a history of events / actualities}\,\,\, -- \,\,\, \textbf{Future} = \textit{a tree of potentialities}
\end{center}
or, put differently, in the \textit{``unbearable'' arrow of time} in the evolution of states of \textit{isolated 
open physical systems}.

In non-relativistic Quantum Mechanics, an \textit{isolated physical system}, $S$, is specified by the following data:
\begin{itemize}
\item[\quad1.]{Physical quantities of $S$ are represented by abstract self-adjoint operators
\[ \widehat{X}=\widehat{X}^{*} \in \mathcal{O}_S, \]
where $\mathcal{O}_S $ is a family of operators whose only properties are that it contains the identity \textbf{1} 
and that if $\widehat{X}\in \mathcal{O}_S$ and $F$ is a real-valued bounded continuous function on the real line then 
$F(\widehat{X}) \in \mathcal{O}_S$. 
It is assumed that, at every time $t$, there is a representation of $\mathcal{O}_S$ on a separable Hilbert space $\mathcal{H}$: 
$\mathcal{O}_{S} \ni \widehat{X} \mapsto X(t),$ where $X(t)$ is a self-adjoint bounded operator on $\mathcal{H}$ (i.e., a
self-adjoint element of $B(\mathcal{H})$).
}
\item[\quad2.]{We specify \,the \textit{quantum\, dynamics}\, of an  \textit{isolated system} \,in \,the \,\textit{Heisenberg
picture:} Operators are time-dependent, their time-dependence is given by the Heisenberg equations
\begin{equation}\label{Heisnberg eq}
 X(t')= e^{i(t'-t)H/\hbar} X(t) e^{-i(t'-t)H/\hbar},\qquad \text{for }\,\, t, t' \,\text{ in } \, \mathbb{R}\,,
 \end{equation}
where $H$ is the Hamiltonian of the system; ($H$ is assumed to be a self-adjoint operator on $\mathcal{H}$). 
Interactions between $S$ and its complement (the rest of the Universe) are assumed to be \textit{negligibly small}. }
\end{itemize}

Because an operator $X$ on $\mathcal{H}$ representing a physical quantity in $\mathcal{O}_S$ is obtained 
by integrating some \textit{``density''} (given by an operator-valued distribution depending on (space and) time, 
such as a \textit{charge-, spin-} or \textit{energy density}) over a compact interval of the time axis, 
every such $X$  is localized in some interval, $I=I_X$, of the time axis. (Bounded functions of) all operators 
localized in an interval $I$ generate a $^{*}$-algebra, $\mathcal{E}_{I}$, and we have that 
$\mathcal{E}_{I'} \subseteqq \mathcal{E}_{I}$ if $I' \subset I$. This allows us to introduce the algebras
\begin{equation}\label{algebra-I}
\mathcal{E}_{\geq t} := \overline{\bigvee_{I \subset [t, \infty)}\mathcal{E}_{I}}\,,\qquad t\in \mathbb{R}\,,
\end{equation}
where the closure is taken in the weak topology of $B(\mathcal{H})$,\footnote{It is convenient to take weak
closures, i.e., consider von Neumann algebras, because they have the property that, with every self-adjoint
bounded operator, they also contain all its spectral projections.} with 
$\mathcal{E}_{\geq t'} \subseteqq \mathcal{E}_{\geq t},$ for $t'>t$.

\vspace{0.15cm}
\noindent \textit{Definition of Events:} A \textit{potential event} that might happen at time $t$ or later
is an element of a partition of unity, 
$\mathfrak{F}:=\big\{\pi_{\xi} \in \mathcal{E}_{\geq t}\, \vert\, \xi \in \mathfrak{X}\big\}$, 
by disjoint orthogonal projections acting on $\mathcal{H}$, where $\mathfrak{X} \,(= $ ``\,spectrum of $\mathfrak{F}$\,'')
is a finite or countably infinite set. The projections $\pi_{\xi} \in \mathfrak{F}$ have the properties

\begin{equation}\label{partition}
  \pi_{\xi}= \pi_{\xi}^{*}, \quad \pi_{\xi}\cdot \pi_{\eta}= \delta_{\xi \eta} \pi_{\xi},\,\,\, \forall\, \xi, \eta \in \mathfrak{X}, \quad 
 \sum_{\xi \in \mathfrak{X}} \pi_{\xi}= \mathbf{1}\,.
\end{equation}
\begin{itemize}
\item[\quad3.]{A \textit{state} of $S$ at time $t$ is given by a \textit{quantum probability measure} 
on the lattice of orthogonal projections in $\mathcal{E}_{\geq t}$, i.e., it is a functional, $\omega_t$, 
with the properties

$\bullet$\, {$\omega_t$ assigns to every orthogonal projection $\pi \in \mathcal{E}_{\geq t}$ a non-negative number 
$\omega_{t}(\pi) \in [0,1]$, with $\omega_{t}(\mathbf{1}) =1$,}

 $\bullet$\, {$\omega_{t}$ is \textit{additive}, in the sense that
\begin{equation} \label{additivity}
\sum_{\pi \in \mathfrak{F}} \omega_{t}(\pi) = 1, \quad \forall\, \text{ partitions of unity }\, 
\mathfrak{F} \subset \mathcal{E}_{\geq t}\,.
\end{equation}}
}
\end{itemize}
\textit{Remark:} \textit{Gleason}'s theorem \cite{G-M} (generalized by \textit{Maeda}) says that states, $\omega_t$, of 
$S$ at time $t$, as defined above, are \textit{positive, normal, normalized linear functionals} on 
$\mathcal{E}_{\geq t}$, i.e., \textit{states} on $\mathcal{E}_{\geq t}$ in the usual sense of this notion.

By definition, 
$$B(\mathcal{H}) \supseteq \mathcal{E}_{\geq t} \supseteq \mathcal{E}_{\geq t'}, \, \,\, \forall t'>t.$$

Thus, apparently, an isolated physical system is characterized by a co-filtration
\[\big\{\mathcal{E}_{\geq t}\,\vert\, t\in \mathbb{R}\big\}\] 
of algebras, with $\mathcal{E}_{\geq t}$ generated by projections representing potential events possibly 
happening at time $t$ or later. 
These data encode all potential events possibly happening in $S$, along with the \textit{deterministic} 
Heisenberg-picture dynamics of $S$. They do \textit{not} contain \textit{any grain of uncertainty or chance, yet}!
This observation compels us to ask where the \textit{probabilistic nature of Quantum Mechanics}, 
expected to be fundamental, is hiding.

It turns out that it is the quantum-mechanical \textit{time evolution of states} of physical systems 
that is \textit{stochastic}. This claim is evident when one considers measurements and observations 
within the so-called \textit{``Copenhagen Interpretation''} of Quantum Me- chanics: Every measurement is thought 
to provoke a non-linear stochastic change of state. An interesting example is the observation of tracks of 
$\alpha$-particles (helium nuclei) emitted from a perfectly round ball of radioactive material in a bubble 
chamber filled with a saturated gas:

\begin{center}
\includegraphics[height=2.9cm]{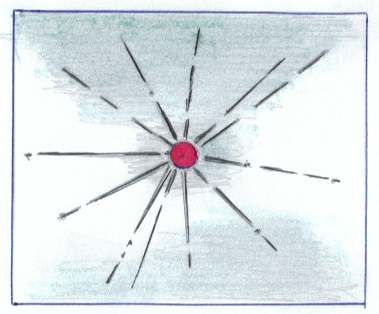}

\vspace{0.1cm}\small{\textit{Figure 3:} Tracks of $\alpha$-particles in a bbubble chamber}
\end{center}
One may wonder how the perfect spherical symmetry of the initial state of this system gets broken in
such a way that tracks of particles pointed in definite directions appear. Physicists tend to agree with
the claim that Quantum Mechanics can \textit{at best} predict the probability distribution on the possible 
directions of such tracks.\footnote{For an explanation of the phenomenon 
of paticle tracks in detectors see \cite{BFF}, and references given there.} But does Quantum Mechanics,
when formulated correctly, explain that tracks (rather than waves) \textit{will} occur; and how do \textit{probabilities} 
enter the theory?

In this section, we summarize a novel approach, dubbed $ETH$-Approach (see \cite{BFSch, FP, Fr}), towards 
understanding the quantum-mechanical \textit{stochastic time evolution of states of isolated open systems} 
featuring events.

\vspace{0.2cm} \textit{Definition 1.} An \textit{Isolated \textbf{Open} Physical System} is an isolated system, $S$, 
characterized by a co-filtration, $\big\{\mathcal{E}_{\geq t}\,\vert\, t\in \mathbb{R}\big\}$, satisfying the 

\vspace{0.2cm}\noindent \textbf{Principle of Diminishing Potentialities ($PDP$):} \textit{In an Isolated Open Physical System 
featuring events the following strict inclusions hold}
\begin{equation}\label{PDP}
\hspace{2.4cm}\mathcal{E}_{\geq t}\,\, \supsetneqq\,\, \mathcal{E}_{\geq t'}\,, \,\text{ for arbitrary }\,\, t'>t\,. \hspace{2cm}
\square
\end{equation}

Note: The ``initial state'' of $S$ may be \textit{pure}; but, since $\mathcal{E}_{\geq t} \subsetneqq B(\mathcal{H})\,, 
\forall t< \infty$, assuming that $(PDP)$ holds, its state at time $t$ will in general be a \textit{mixed} state on 
$\mathcal{E}_{\geq t}$ (\textit{entanglement!})
This observation opens the door towards a natural notion of \textit{Actual Events}, or
\textit{``actualities''},  in our formalism and to a theory of direct\,/\,projective 
mea- surements and observations.

In accordance with the ``Copenhagen interpretation'' of Quantum Mechanics, one may pretend that some 
\textit{potential event} in a partition of unity $\mathfrak{F} =\lbrace \pi_{\xi} \vert \xi \in \mathfrak{X} \rbrace 
\subset \mathcal{E}_{\geq t}$ \textit{actually happens} in the interval $[t, \infty)$ of times, i.e., becomes an 
\textit{actual event} setting in at time $t$, iff
\begin{equation}\label{incoherent sp}
\omega_{t}(A) = \sum_{\xi \in \mathfrak{X}} \omega_{t}(\pi_{\xi} \, A\, \pi_{\xi}),\quad \forall A \in \mathcal{E}_{\geq t}\,.
\end{equation}
No off-diagonal elements appear on the right side of \eqref{incoherent sp}, which describes an \textit{incoherent} 
superposition of states in the images of disjoint projections.

But what, in Quantum Mechanics, determines whether, at some time $t$, a partition of unity $\mathfrak{F}$ 
exists such that Eq.~\eqref{incoherent sp} holds for the state of the system at time $t$? This is the question 
to be answered next. To answer it one would clearly \textit{not} want to invoke anything like the
\textit{``free will''} of observers! 

In order to describe what I consider to be a satisfactory answer, I have to introduce 
some simple notions from algebra.\footnote{The following remarks may look overly abstract; but this cannot 
be avoided, because the algebras $\mathcal{E}_{\geq t}$ are usually not of type $I$.} Let $\mathfrak{M}$ 
be a $^{*}$-algebra, and let $\omega$ be a state on $\mathfrak{M}$. We define the \textit{centralizer} of
$\omega$ by
\[\mathcal{C}_{\omega}(\mathfrak{M}):= \big\{ X \in \mathfrak{M}\, \big|\, \omega([A,X])=0,\,\, \forall A\in \mathfrak{M} \big\}\]
Note that $\mathcal{C}_{\omega}(\mathfrak{M})$ is a subalgebra of $\mathfrak{M}$ and that $\omega$ 
is a normalized trace on $\mathcal{C}_{\omega}(\mathfrak{M})$. The 
\textit{center}, $\mathcal{Z}_{\omega}(\mathfrak{M})$, of $\mathcal{C}_{\omega}(\mathfrak{M})$ is defined by
\begin{equation}\label{center}
\mathcal{Z}_{\omega}(\mathfrak{M}):= \big\{ X\in \mathcal{C}_{\omega}(\mathfrak{M}) \,\big| \, [X, A] =0,\,\forall 
A \in \mathcal{C}_{\omega}(\mathfrak{M}) \big\}
\end{equation}

Let $S$ be an isolated open physical system (see Definition 1, above). In \eqref{center} we set 
$\mathfrak{M}:=\mathcal{E}_{\geq t}$, and $\omega :=\overline{\omega}_t$, where $\overline{\omega}_t$ is a state
on $\mathcal{E}_{\geq t}$.

\textit{Definition 2.} Given that $\overline{\omega}_{t}$ is some state on $\mathcal{E}_{\geq t}$, we say that an 
\textit{Actual Event} is setting in at time $t$ iff $\mathcal{Z}_{\overline{\omega}_{t}}(\mathcal{E}_{\geq t})$ 
contains \textit{at least} two non-zero orthogonal projections, $\pi^{(1)}, \pi^{(2)}$, which are disjoint, 
i.e., $\pi^{(1)}\cdot \pi^{(2)} =0$, and have non-vanishing \textit{``Born probabilities''}, i.e.,
\[\hspace{2.4cm}0< \overline{\omega}_{t}(\pi^{(i)})< 1\,, \quad \text{ for  }\, i=1,2\,.\hspace{2.5cm}\square\]
Let us suppose for simplicity that $\mathcal{Z}_{\overline{\omega}_{t}}(\mathcal{E}_{\geq t})$ is generated 
by a partition of unity $\mathfrak{F}_t =\lbrace \pi_{\xi}\,\vert\, \xi \in \mathfrak{X}_{\overline{\omega}_t}\rbrace$ of 
orthogonal projections, where $\mathfrak{X}_{\overline{\omega}_t}$, the \textit{spectrum} of the abelian algebra 
$\mathcal{Z}_{\overline{\omega}_{t}}(\mathcal{E}_{\geq t})$, 
is assumed to be a \textit{countable} set. Then Eq.~\eqref{incoherent sp} holds true for $\mathfrak{X}=\mathfrak{X}_{\overline{\omega}_t}$.

 The \textbf{Law} describing the stochastic time evolution of states in Quantum Mechanics is derived from the following 
 \textit{State Reduction-, or Collapse Postulate} (which makes precise mathematical sense if \textit{time} is \textit{discrete}).
 
 Let $\omega_t$ be the state of $S$ at time $t$. Let $dt$ denote a time step; ($dt$ is strictly positive if time is discrete; 
 otherwise we will attempt to let $dt$ tend to 0 at the end of our constructions). We define a state 
 $\overline{\omega}_{t+dt}$ on the algebra $\mathcal{E}_{\geq t+ dt} (\subset \mathcal{E}_{\geq t})$ by 
 restriction of $\omega_t$, i.e.,
 \[\overline{\omega}_{t+dt}:= \omega_{t}\big|_{\mathcal{E}_{\geq t+dt}}\,.\]

\noindent {\textbf{Axiom CP:}} \,\,\textit{Let $\mathfrak{F}_{t+dt}:= \big\{\pi_{\xi}\,|\,\xi\in \mathfrak{X}_{\overline{\omega}_{t+dt}}\big\}$ be the 
partition of unity generating the center, $\mathcal{Z}_{\overline{\omega}_{t+dt}}(\mathcal{E}_{\geq t+dt}),$
of the centralizer of the state $\overline{\omega}_{t+dt}$ on $\mathcal{E}_{\geq t+dt}$.
Then \textit{`Nature'} replaces the state $\overline{\omega}_{t+dt}$ by a state}
\[ \omega_{t+dt}(\cdot) \equiv \omega_{t+dt, \xi}(\cdot):=
\overline{\omega}_{t+dt}(\pi_{\xi})^{-1} \cdot \overline{\omega}_{t+dt}(\pi_{\xi}(\cdot)\pi_{\xi})\,,\]
\textit{for some $\xi \in \mathfrak{X}_{\overline{\omega}_{t+dt}}$, with} $\overline{\omega}_{t+dt}(\pi_{\xi})\not= 0$.

\textit{The probability,} $\text{prob}_{t+dt}(\xi)$, \textit{for the state $\omega_{t+dt,\xi}$ to be selected by \textit{`Nature'} 
as the state of $S$ at time $t+dt$ is given by}
\begin{equation}\label{Born Rule}
\hspace{2.2cm} \text{prob}_{t+dt}(\xi)= \overline{\omega}_{t+dt}(\pi_{\xi}) \quad \quad \text{(Born's Rule)}
\hspace{1.7cm}\square
\end{equation}

The $ETH$-Approach to Quantum Mechanics sketched above leads to the 
following picture of dynamics in Quantum Mechanics: 
The \textit{evolution of states} of an isolated open system $S$ featuring events, in the sense of \textit{Definition 2}, 
is determined by a \textit{stochastic branching process}, whose state space is referred to 
as the \textit{non-commutative spectrum}, $\mathfrak{Z}_{S}$, of $S$ (see~\cite{Fr}). Assuming 
that all the algebras $\mathcal{E}_{\geq t}$ are isomorphic to one specific (universal) von Neumann 
algebra, denoted by $\mathcal{N}$,\footnote{This is the case  in relativistic Quantum Electrodynmics  \cite{Buchholz}} the non-commutative spectrum of $S$ is defined by
\begin{equation}\label{NCspect}
\mathfrak{Z}_{S}:= \bigcup_{\omega} \Big(\,\omega\,, \mathcal{Z}_{\omega}(\mathcal{N})\Big)\,, 
\end{equation}
where the union over $\omega$ is a disjoint union, and $\omega$ ranges over \textit{all} states of $S$ of physical interest.\footnote{``States of physical interest'' are normal states a concrete system can actually be prepared in. Here we may
leave this notion a little vague.}
If time is taken to be discrete then \textit{Definition 2} of actual events and Born's Rule \eqref{Born Rule} 
completely specify the \textit{branching probabilities} of this process. The mathematical theory obtained 
when the time step, $dt$, tends to 0 is, however, not developed rigorously, yet. I expect that this is will 
become a challenging topic for mathematicians.

\vspace{0.2cm}\textit{Remarks:}
\begin{itemize}
\item{Here is an explanation of the meaning of the name ``$ETH$-Approach'': ``$E$'' stands for ``events'', 
``$T$'' for ``trees'' -- referring to the tree-like structure of the space of all actualities an isolated physical system 
could in principle encounter in the course of its evolution -- and ``H'' stands for ``histories,'' 
referring to the \textit{actual trajectory} of states occupied by the system in the course of 
its evolution.}

\item{{\textbf{Axiom CP}} (the Collapse Postulate), in combination with Eq.~\eqref{incoherent sp} and Born's Rule 
\eqref{Born Rule}, is reminiscent of the collapse postulate of the Copenhagen interpretation of Quantum
Mechanics. But the Principle of Diminishing Potentialities $(PDP)$ imparts on it a transparent, logically 
coherent status, without assigning any role to something like the free will of ``observers.''}

\item{One might argue that  $(PDP)$ and the Collapse Postulate just represent a precise version of the 
\textit{Many-Worlds Interpretation} of Quantum Mechanics \cite{Everett}. However, the $ETH$-Approach provides
a precise rule for \textit{``branching,''} namely \textbf{Axiom CP} along with Born's Rule -- which the 
Many-Worlds interpretation does \textit{not}. And in the $ETH$-Approach there is no reason to imagine that, 
besides the world we actually perceive, other worlds exist. 
}

\item{In \cite{Fr} it is argued in which way the occurrence of an actual event may correspond 
to the \textit{measurement of a physical quantity.} (Of course, in general, there are plenty of actualities happening 
that cannot be related to the measurement of a well-defined physical quantity.) Let $\varepsilon \ll 1$ 
be a positive number. We define $\mathcal{Z}_{t, \varepsilon}$ to be the abelian algebra \
generated by the projections in a family, 
$\big\{\pi_1, \dots, \pi_n\big\} \subset \mathcal{Z}_{\overline{\omega}_t}(\mathcal{E}_{\geq t}),$
 of $n$ disjoint orthogonal projections with the property that
$$\overline{\omega}_{t}(\mathbf{1}-\sum_{i=1}^{n}\pi_i)\leq \varepsilon$$
Let $X=X^{*}$ be a self-adjoint operator in the algebra $\mathcal{E}_{\geq t}$ representing some physical
quantity $\widehat{X} \in \mathcal{O}_S$, and let $\big\{\Pi_1, \dots, \Pi_k\big\}$ be $k$ disjoint spectral projections of $X$
with the property that $\overline{\omega}_{t}(\mathbf{1}- \sum_{i=1}^{k} \Pi_i)\leq \varepsilon$. 
If every projection $\Pi_{i}, i=1, \dots, k,$ can be approximated in norm by operators in $\mathcal{Z}_{t, \varepsilon}$ 
with a precision of $\mathcal{O}(\varepsilon)$ then one can argue that the occurrence of an event in 
$\mathcal{Z}_{\overline{\omega}_t}(\mathcal{E}_{\geq t})$
corresponds to the measurement of an approximate value of the physical quantity $\widehat{X}$ at a time 
$\approx t$, with a precision of $\mathcal{O}(\varepsilon)$.

A more compelling discussion of the relationship between actual events and measurements of physical quantities
can be found, in the context of an analysis of simple models, in Sect.~5.4 of \cite{FP} and will be pursued in
forthcoming work.
} 
\item{For theoretical physicists, the most interesting problem connected with the $ETH$-Approach 
to Quantum Mechanics is to explore physical mechanisms leading to the Principle of Diminishing Potentialities 
$(PDP)$. Some simple models satisfying this principle are discussed in Sect.~5 of \cite{FP}. 
These models have the feature that if time is continuous their total Hamiltonian, $H$, is unbounded 
above \textit{and} below. I have argued that $(PDP)$ is compatible with the usual 
spectrum condition, namely that $H\geq 0$, \textit{only} in models of \textit{local relativistic quantum theory} -- a rather
tantalizing conclusion. In such models, a mechanism leading to $(PDP)$ is known: It is related to 
\textit{Huygens' Principle} valid in quantum electrodynamics and other theories with massless particles 
in four-dimensional space-time; see \cite{Buchholz}.

We expect that there are plenty of other mechanisms leading to $(PDP)$; but a systematic study has not been
made, yet.

A variant of the $ETH$-Approach formulated within local relativistic quantum theory is discussed in \cite{Fr}. 
}
\end{itemize}

\section{Conclusions and Acknowledgements}\label{UCA}

It is a widespread conviction that the fundamental Laws of Nature treat past and future in a symmetric way 
and that, for this reason, they cannot only be used to predict the future, but they can also be applied to reconstructing
the past from knowledge of the present. Irreversible behavior of physical systems and an arrow of time 
are thus often perceived as mysterious. In Sections 3 through 5, I have examined examples of physical
systems exhibiting an \textit{arrow of time} or \textit{irreversible behavior} in spite of the fact that the underlying 
dynamics of these systems is \textit{time-reversal invariant}. In this context, results of Elliott Lieb and coworkers 
concerning properties of quantum-mechanical entropy furnish some of the elements of an explanation.

In Sect.~6, I have argued, I hope quite convincingly, that the quantum-mechanical evolution of states of 
physical systems featuring events that can in principle be obser- ved is fundamentally \textit{stochastic} 
and exhibits an \textit{arrow of time.} 
So far, this view has not been widely accepted, yet.\footnote{To quote Max Planck: \textit{``Eine neue 
wissenschaftliche Wahrheit pflegt sich nicht in der Weise durchzusetzen, daß ihre Gegner überzeugt 
werden und sich als belehrt erklären, sondern vielmehr dadurch, daß ihre Gegner allmählich aussterben 
und daß die heranwachsende Generation von vornherein mit der Wahrheit vertraut gemacht ist.''}} 
However, I am confident that it has its merits and that it is firmly rooted in known properties of 
quantum field theories with massless modes, such as quantum electrodynamics. Moreover, 
simple models implement and illustrate it in a natural and convincing way (see \cite{FP}).

One might argue that the most fascinating example of a physical system exhibiting an arrow of time is the 
\textit{Universe.} The ultimate theory to be used to describe the Universe is expected to be some 
generalization of quantum mechanics that people tend to call ``quantum gravity.'' 
This theory, whose precise formulation remains to be discovered, ought to describe the big bang and 
the ensuing evolution of the Universe. Fundamental problems in a search for ``quantum gravity'' are, 
among others, to introduce a sharp and useful concept of \textit{cosmological ``event''} and to formulate a 
precise \textit{stochastic law of evolution} of \textbf{``space-time-matter.''} I expect that the ideas sketched 
in Sect.~6 and the results described in \cite{Fr} might become relevant in this connection. 
A future theory should then elucidate why and how \textit{structure with classical features} 
emerged from an initial highly isotropic and homogeneous quantum state of the Universe 
(during an epoch when there weren't any ``observers'' to perform measurements).

Of course, I am not competent to do justice to this example. Nevertheless I would like to sketch some 
\textit{basic facts} related to the fundamentally \textit{irreversible evolution of the state of the Universe} 
that remain highly enigmatic and wait to be unravelled.

\begin{itemize}
\item{On very large distance scales, the Universe appears to be isotropic and homogeneous, 
although different patches could not have been causally connected at very early times. It has been proposed
that \textit{Inflation}, an epoch of glaringly irreversible, explosive evolution of the Universe, could explain 
these surprising features (besides solving several other puzzles).}
\item{The Universe \textit{expands} (rather than contracts), and its energy density, $\rho$, appears
to be very nearly equal to the critical density for an open universe. Inflation appears to
provide an explanation of this.}
\item{The Universe exhibits a basic asymmetry between Matter and Antimatter.}
\item{Galactic rotation curves and plenty of other phenomena (including, e.g., gravitational lensing)
suggest that there exists Dark Matter in the Universe with an equation of state implying that 
$0<p \ll \rho$, where $p$ denotes pressure.}
\item{The observed \textit{accelerated expansion} of the Universe suggests that there exists 
Dark Energy with an equation of state $p\approx - \rho$.}
\item{At the present stage of evolution of the Universe, the contributions of visible matter, Dark Matter and
Dark Energy to the energy density of the Universe have the same order of magnitude.}
\item{There exist tiny, highly homogeneous intergalactic magnetic fields in the Universe stretched out
over huge distances.\\
...}
\end{itemize}
A \textit{daring conjecture:} The facts just described must have something 
to do with one another and ought to have interrelated theoretical explanations! One may safely expect that these
explanations will be based, in part, on ``physics beyond the standard model'' (e.g., they are likely to involve  
quantum fields of geometrical nature not present in the standard model of particle physics that will give rise 
to Dark Matter and Dark Energy). They may thus lead theorists to discover genuinely new physics.\footnote{The reader 
may find a sketch of my views concerning these matters in \cite{Fr, Froh} and references given there. 
Not being closely familiar with this subject I must refrain from quoting any literature.}\\

To conclude, one may hope that revealing origins of \textit{time's arrow} and of \textit{irreversibility} 
in different physical theories remains a challenging endeavor in theoretical physics that is likely to give 
rise to surprising future discoveries.

\textbf{Acknowledgements.}
I wish to thank my collaborators, including W.K.~Abou Salem, V.~Bach, Ph.~Blanchard, W.~De Roeck, M.~Griesemer,
A.~Knowles, M.~Merkli, A.~Pizzo, B.~Schlein, B.~Schubnel, I.M.~Sigal, D.~Ueltschi, Zhou Gang, and others, 
for the joy of many quite successful joint efforts on projects related to material reviewed in this paper.

I am grateful to these as well as to many further colleagues and friends, including D.~Brydges, D.~Buchholz, 
the late D.~D\"urr, S.~Goldstein, G.M.~Graf, E.~Seiler and T.~Spencer, for countless useful discussions 
on some of the ideas described in this paper. I thank T.~Spencer for hospitality at the School of 
Mathematics of the Institute for Advanced Study, during a period when I was working on the material 
sketched in Section 6, and Ph.~Blanchard and H.~Siedentop for hospitality at their respective 
institutions.

I am deeply thankful to Elliott Lieb for all he has taught me and for his friendship, and to my 
PhD advisor and colleague Klaus Hepp for -- among many other things -- having introduced 
me to Elliott and for sharing him with me as a common friend.

\end{document}